\documentclass[aps,nofootinbib,superscriptaddress, showpacs,preprintnumbers,  nofootinbibt,twocolumn]{revtex4-2}
\usepackage{epsfig}
\usepackage{multirow}
\usepackage{eurosym}
\usepackage{dcolumn}
\usepackage{bm}
\usepackage{enumerate}
\usepackage{float}
\usepackage{epstopdf}
\usepackage{amsmath}
\usepackage{bm}
\usepackage{amsfonts}
\usepackage{amssymb}
\usepackage{graphicx}
\usepackage{alphalph,mathtools}
\usepackage{etoolbox}
\usepackage{color}
\usepackage{booktabs}
\usepackage{footnote}
\usepackage{makecell,tabularx}
\usepackage{hyperref}
\hypersetup{colorlinks,citecolor=blue}
\hypersetup{colorlinks=true,linkcolor=red,filecolor=magenta,urlcolor=blue}

\def\be{\begin{equation}}
    \def\ee{\end{equation}}
\def\bea{\begin{eqnarray}}
    \def\eea{\end{eqnarray}}

\begin{document}

\title{Diagnostic and Comparative Analysis of Dark Energy Models with $q(z)$ Parametrizations}

\author{Dhruv Arora}
\email{arora09dhruv@gmail.com}
\affiliation{Pacif Institute of Cosmology and Selfology (PICS), Sagara, Sambalpur 768224, Odisha, India}
\author{Himanshu Chaudhary}
\email{himanshuch1729@gmail.com} \affiliation{Department of Applied Mathematics, Delhi Technological University, Delhi-110042, India}
\affiliation{Pacif Institute of Cosmology and Selfology (PICS), Sagara, Sambalpur 768224, Odisha, India}
\author{Shibesh Kumar Jas Pacif}
\email{shibesh.math@gmail.com}
\affiliation{Pacif Institute of Cosmology and Selfology (PICS), Sagara, Sambalpur 768224, Odisha, India}
\author{G.Mustafa}
\email{gmustafa3828@gmail.com} \affiliation{Department of Physics,
Zhejiang Normal University, Jinhua 321004, Peoples Republic of
China}
\affiliation{New Uzbekistan University, Mustaqillik ave. 54, 100007 Tashkent, Uzbekistan}

\begin{abstract}
This manuscript presents a diagnostic analysis of three dark energy models resulting from the parametrization of the deceleration parameter. These models exhibit intriguing features, including late-time acceleration and a cosmological phase transition from early deceleration to late acceleration. The analysis utilizes parametrizations of the deceleration parameter, $q(z)$, and employs Cosmic Chronometers (CC), Type Ia supernovae (SNIa), Gamma Ray Bursts (GRB), Quasar (Q) and Baryon Acoustic Oscillations (BAO) datasets to constrain the models and determine the best-fitting values of the model parameters. Additionally, the evolution of kinematic cosmographic parameters is investigated. The study focuses on discussing the statefinder and Om diagnostic analyses of the considered models, comparing them with the well-established $\Lambda$CDM and SCDM models. By utilizing information criteria, the viability of the models is examined, assessing their goodness of fit and their ability to explain the observed data. The results provide valuable insights into the behavior and characteristics of the dark energy models. The comparison with the standard models sheds light on the similarities and differences, while the information criteria analysis offers a quantitative assessment of their suitability. This analysis contributes to our understanding of the dynamics and evolution of the Universe, furthering our knowledge of dark energy and its role in shaping the cosmos.
\end{abstract}

\maketitle
\tableofcontents

\section{Introduction}\label{sec1}
The discovery of late-time cosmic acceleration\cite{1} established a new area of study in modern cosmology. To explain this accelerated phenomenon, several theoretical models have been developed. The majority of them are based on either a modification of the Einstein-Hilbert action \cite{2} or the existence of new types of exotic fields in nature, dubbed "dark energy" (DE). This paper will concentrate on the second aspect and consider DE as the driving agent for the universe's current accelerated expansion, which is viewed as a hypothetical energy component with high negative pressure. Several DE models have been investigated in the last two decades to account for this phenomenon (for review, see references \cite{3,4}. Despite these efforts, however, the true nature of dark energy remains unknown. The $\Lambda$CDM model is the most popular and simplest cosmological DE model and agrees well with recent observational data. The $\Lambda$CDM model introduces the famous cosmological constant into general relativity and solves the equation of state parameter $\omega =-1$. However, it has two significant flaws: fine-tuning and cosmological coincidence issues \cite{5,6}. This motivates theorists to develop alternative DE models, such as quintessence, Chaplygin gas, Chameleon, etc.\\\\
Due to their dynamical nature, models are widely used in addressing several issues in cosmology \cite{7,8,9,10,11,12}, in which the equation of state parameter evolves dynamically with time, in contrast to cosmological constant models. In order to generate sufficient negative pressure and acceleration, the potential term for the quintessence field usually dominates over its kinetic part. To that end, a wide range of quintessence potentials has been considered.
However, none of these models has much empirical support. The reviews on DE models \cite{13,14} contain some excellent descriptions in this area. According to theory, without DE, the Universe should slow down because gravity holds matter together. The gravitational instability theory of structure formation and big bang nucleosynthesis both support the existence of an early decelerated expansion phase of the Universe. In order to incorporate both scenarios, the deceleration parameter must exhibit a signature flip. To put it another way, a transition from a decelerating phase ($q > 0$) to a late-time accelerating phase ($q< 0$) is required to explain both structure formation and current acceleration measurements.\\\\
Recently, various studies have been performed to analyze the kinematics of the Universe through phenomenological parametrizations of $q(z)$ and $\omega(z)$ \cite{EoS1,EoS2,EoS3,kundu2024gravitational,q(z)1,q(z)2,q(z)3,q(z)4,q(z)5,q(z)6,q(z)7,q(z)8,q(z)9,q(z)10,q(z)11,q(z)12,q(z)13}. The basic characteristics of dynamical evolution, both static and dynamic, can be expressed in terms of the Hubble parameter $H$ and the deceleration parameter $q$ only. In fact, these two parameters enable us to construct model-independent kinematics of the cosmological expansion \cite{17}. Usually, the kinematic approach is described by a particular metric theory of gravity. It does not depend on the validity of general relativity or any model-specific assumptions like the matter-energy content of the Universe. Moreover, the analysis of nonstandard cosmological models, as explored by \cite{doroshkevich1989large}, also presents a novel perspective on cosmological dynamics. These models incorporate unstable dark matter alongside a cosmological term, yielding a time-dependent dark energy model. This leads to dynamic changes in the behavior of dark energy over cosmic time scales and offers a unique framework for understanding the intricate interplay between different components of the universe and provides valuable insights into the evolution of cosmic structures and phenomena. Motivated by these facts, we considered some well-motivated two-parameter parametric forms of $q(z)$ to consider here. We have found some observational constraints on the model parameters. The evolutionary behavior of DE and the Universe's expansion history for these parameterizations are investigated. The properties of parametrizations are discussed in Section \ref{sec2}. Section \ref{sec3} examines the constraints on the various reconstructed model parameters using observational datasets of Cosmic Chronometers (CC), Type Ia supernovae (SNIa), Gamma Ray Bursts (GRB), Quasar (Q) and Baryon Acoustic Oscillations (BAO). In Section \ref{sec4}, we compare the theoretical and observational Hubble functions of all the models. Section \ref{sec5} gives detailed treatment of the cosmographic parameters (deceleration and jerk) and presents the comparison with the $\Lambda$CDM model. Section \ref{sec6} discusses the state finder diagnostics, which is a technique generally used to distinguish various dark energy models and compare their behavior using the higher-order derivatives of the scale factor. Section \ref{sec7} introduces the Om diagnostics which is another method to compare different dark energy models using slope variation. Section \ref{sec8} gives a brief about information criteria. Results and conclusions are discussed in section \ref{sec9} and \ref{sec10} respectively. 

\section{Parameterization of $q(z)$ and the corresponding models.}\label{sec2}
Cosmography is an approach used to study the Universe's expansion history, which is purely based on the cosmological principle, stating that the Universe is homogeneous and isotropic on large scales \cite{18}. Basic properties of the Universe can be expressed in terms of the Hubble parameter ($H_0$) and the deceleration parameter ($q_0$). These parameters enable us to construct model-independent kinematics of cosmological expansion. In the standard approximation, the scale factor $a(t)$ can be expanded as a Taylor expansion around the present time $t_0$ (which is also the current age of the Universe) \cite{pacif2020dark}. This can be written as:
\begin{equation}
	\begin{split}
		a^{(n)} = 1+H_0(t-t_0)-\frac{1}{2!}q_0H_0^2(t-t_0)^2+\\\frac{1}{3!}j_0H_0^3(t-t_0)^3+\frac{1}{4!}s_0H_0^4(t-t_0)^4+\frac{1}{5!}l_0H_0^5(t-t_0)^5  
	\end{split}
 \label{1}
\end{equation}
where $H_0$ is the Hubble parameter measuring velocity, $q_0$ is the deceleration parameter measuring acceleration, $j_0$ is the jerk parameter measuring change in acceleration, $s_0$ is the snap parameter and $l_0$ is the lerk parameter. All these kinematic variables play a major role in the cosmographic analysis of the Universe and help distinguish various dark energy models. The evolution of the deceleration parameter can be described in terms of cosmic parameters such as the scale factor $a$, the cosmic redshift $z$, and time $t$. However, since the exact functional form of the deceleration parameter is unknown, we make use of Taylor expansion of $q$ in terms of a general variable $x$ \cite{20}:
\begin{eqnarray}
	q(x) = q_0+q_1(1-\frac{x}{x_0})+q_2(1-\frac{x}{x_0})^2+....
 \label{2}
\end{eqnarray}
It is known that if the number of terms in the expansion is increased, we can get a better fit of the observational data, but to preserve the simplicity of the model, we only take the first two terms in the expansion. In this paper, three parametrizations \cite{pacif2020dark} of the deceleration parameter are considered as a function of cosmological redshift $z$ in the forms:
\begin{equation}
	q(z) = q_0+q_1z
\end{equation}
\begin{equation}
	q(z)= q_0+\frac{q_1z}{1+z}
\end{equation}
\begin{equation}
	q(z) = q_0+\frac{q_1z(1+z)}{1+z^2},
\end{equation}
where $z= \frac{a_0}{a}-1$ is the cosmic redshift and $q_0$ is the present value of deceleration parameter. According to recent cosmological observations, the Universe is accelerating. This means that the Universe must have passed through a period of slower expansion phase \cite{21}. Also, a decelerating phase is crucial for large-scale structure formation. This phase transition from decelerating to accelerating expansion is highly important when describing the Universe's dynamics. The present cosmic behavior can be estimated by taking the present value of decelerating parameter in the negative domain. Also, in all the linearly varying deceleration parameters, the $q_1$ must be positive for the uUiverse to transition from deceleration to acceleration. For this study, we have considered the present value of $q_0$ and the value of $q$ when $z=1$ to be within the $1\sigma$ limit from the observational data. In order to find the constraints on the cosmic parameters, we make use of 31 points of the Cosmic chronometers dataset, 1048 points of the supernova type Ia, 17 points of BAO and 162 points of the GRB dataset and 24 observations of the compact radio quasar obtained from a VLBI all-sky survey of 613 milliarcsecond ultra-compact radio sources at $\mathrm{GHz}$

\subsection{Model 1}

The deceleration parameter $q = -\Ddot{a}a/\dot{a}^2$ \cite{22} can be rewritten in the form,
\begin{eqnarray}
	q(z) = \frac{d(lnH)}{dz} (1+z) -1
 \label{6}
\end{eqnarray}
where an overhead dot represents the derivative w.r.t time. Proceeding further, we present our analysis of the first parametrization of $q(z)$, which takes the functional form as:
\begin{eqnarray}
	q(z)=q_0+q_1z
 \label{7}
\end{eqnarray}
where the factor of $(\frac{x}{x_0})$ is replaced by $a_0/a = 1+z$. Moving further, from Eq. (\ref{6}), we can also find the expression for the Hubble parameter as follows:
\begin{eqnarray}
	H(z)= H_0(1+z)^{1+q_0-q_1} e^{q_1z}
\end{eqnarray}
The accelerating redshift \cite{22} $(q(z_{acc})=0)$ for the parametrization is given by:
\begin{eqnarray}
	z= \frac{-q_0}{q_1}
\end{eqnarray}

\subsection{Model 2}
Due to the constant growth of $q$, some authors suggested a new functional form of deceleration parameter, which is famously known as the Chevallier-Polarski-Linder (CPL) parametrization \cite{23}. This can be obtained by substituting $\frac{x}{x_0} = \frac{a}{a_0}$ in Eq. (\ref{2}). This is given by:
\begin{eqnarray}
	q(z) = q_0+\frac{q_1z}{1+z}
\end{eqnarray}
The value of $q$ in the infinite past is given by $q(\infty) = q_0+q_1$. The accelerating redshift and the expression for Hubble parameter \cite{22} for this function is given by:
\begin{eqnarray}
	H(z) = H_0(1+z)^{1+q_0-q_1}e^{\frac{-q_1z}{1+z}}
\end{eqnarray}
\begin{eqnarray}
	z=-(1+\frac{q_1}{q_0})^{-1}
\end{eqnarray}

\subsection{Model 3}

The next parametrization of $q(z)$ is inspired by a very famous divergence-free parametrization of dark energy equation of state \cite{21}, which is given as:
\begin{eqnarray}
	q(z) = q_0+q_1z\frac{1+z}{1+z^2}
\end{eqnarray}
This is a much more suitable form of deceleration parameter and can be obtained by substituting $\frac{x}{x_0} = 1-\frac{z(1+z)}{1+z^2}$
in Eq. (\ref{2}). The following equations can obtain the accelerating redshift and the expression for the Hubble parameter:

\begin{eqnarray}
	H(z) = H_0(1+z)^{1+q_0} (1+z^2)^{\frac{q_1}{2}}
\end{eqnarray}
\begin{eqnarray}
	z = \frac{-(1+\frac{q_0}{q_1}) \pm \sqrt{(1+\frac{q_0}{q_1})^{-2}-4(1+\frac{q_0}{q_1})^{-1}}}{2}
\end{eqnarray}

\section{Data Analysis}\label{sec3}
To thoroughly examine the characteristics of the considered models, it is crucial to determine the optimal values for the model parameters, namely $q_0$ and $q_1$, in conjunction with the present day Hubble function $H_0$. Our analysis employs three observational datasets: the cosmic chronometers (CC) dataset, Type Ia Supernova dataset, and the Baryon Acoustic Oscillations (BAO) dataset, which consists of 31, 1048, and 17 data points, respectively. To constrain the parameters of the cosmological models, we employ a robust statistical approach combining the standard Bayesian technique, likelihood function methodology, and the Markov Chain Monte Carlo (MCMC) method. This comprehensive approach ensures reliable and accurate estimation of the model parameters based on observational data.

\subsection{Methodology}
Markov Chain Monte Carlo (MCMC) is a powerful statistical technique widely used in cosmology to explore and constrain the parameter space of cosmological models. It provides a robust and efficient way to estimate and quantify the uncertainties of model parameters based on observational data \cite{gelman2013bayesian}. MCMC algorithms are particularly valuable in cosmology for several reasons. Firstly, they allow for Bayesian inference, incorporating both prior knowledge and observational data to generate posterior probability distributions that represent parameter uncertainties. This is crucial in cosmology, where parameters are interconnected and their interdependencies need to be properly accounted for. Another advantage of MCMC is its ability to efficiently explore complex parameter spaces. Cosmological models often involve a large number of parameters, and exhaustive grid-based searches become computationally challenging. MCMC overcomes this limitation by adaptively sampling regions of higher posterior probability, making it well-suited for cosmological parameter estimation \cite{lewis2002cosmological,gelman2013bayesian} MCMC also provides insights into parameter degeneracies and correlations. Cosmological models can exhibit degeneracies, where different combinations of parameters produce similar observational predictions. MCMC algorithms capture these degeneracies by sampling the joint posterior distribution, revealing parameter correlations and aiding in the interpretation of observational constraints \cite{tegmark1997karhunen}. To use MCMC in cosmology, a likelihood function is defined to quantify the agreement between model predictions and observed data. This likelihood, combined with prior knowledge, forms the posterior distribution. MCMC algorithms, such as the Metropolis-Hastings algorithm or the Gibbs sampler, sample from this posterior distribution, generating a chain of parameter samples that converge to the true posterior. Careful implementation of MCMC is essential for reliable cosmological analyses. Convergence diagnostics, such as the Gelman-Rubin statistic, ensure that the algorithm has adequately explored the parameter space. Attention should be given to the choice of proposal distributions and step sizes, as they can affect the efficiency and convergence properties of the algorithm.

\subsection{Data Description}

\subsubsection{Cosmic Chronometers (CC) Dataset}
To constrain the parameters of each model in our analysis, we utilize the Cosmic Chronometers (CC) dataset, which is derived from measurements of the Baryon Acoustic Oscillations (BAO) in the radial direction of galaxy clustering \cite{H(z)}. The CC dataset can also be obtained through the differential age approach, where the redshift dependence of the Hubble function is determined from the derivative of redshift with respect to time for two passively evolving galaxies. In our study, we incorporate 31 data points for CC spanning the redshift range $0.07 \leqslant z \leqslant 2.42$. To estimate the model parameters $H_0$, $q_0$ and $q_1$, We compare the theoretical predictions of each model with the observational data using the chi-square function. The chi-square function is defined as:
\begin{equation}
\chi_{CC}^2(H_0, q_0, q_1) = \sum_{i=1}^{31} \frac{\left[H_{\text{th}}(z_i, H_0, q_0, q_1) - H_{\text{obs}}(z_i)\right]^2}{\sigma_{H(z_i)}^2},
\end{equation}
where $H_{\text{th}}$, $H_{\text{obs}}$, and $\sigma_{H(z_i)}$ denote the model's predicted Hubble rate, the observed value of the Hubble rate, and the standard error at redshift $z_i$, respectively. The numerical values of the Hubble function at the corresponding redshifts are presented in \cite{niu2023cosmological}. This analysis allows us to determine the best-fit parameters for each model by minimizing the chi-square function and assessing the level of agreement between the models and the observed data.
\subsubsection{Type Ia Supernova Datasets}
The observation of Type Ia supernovae (SNIa) has played a pivotal role in the discovery of cosmic acceleration, shedding light on the nature of dark energy. Over the years, several compilations of SNIa data have been released  \cite{Pan1,Pan2,Pan3,Pan4,Pan5} ,Taking into account the measurements of the Pantheon Type Ia supernova dataset \cite{scolnic2018complete} and observations of quasars \cite{quasers} and gamma-ray bursts (GRBs) \cite{GRB}, the model parameters can be constrained by comparing the observed distance moduli, denoted as $\mu_i^{\text{obs}}$, with the theoretical distance moduli, denoted as $\mu_i^{\text{th}}$. The distance modulus is defined as $\mu = m - M$, where $m$ and $M$ represent the apparent and absolute magnitudes, respectively. The term $\mu_0 = 5\log_{10}(H_0^{-1}/\text{Mpc}) + 25$ is a nuisance parameter that can be marginalized over. The luminosity distance $D_L(z)$ is given by the integral expression:
\begin{equation}
    D_L(z) = \frac{c}{H_0}(1+z) \int_0^z \frac{dz^*}{E(z^*)},
\end{equation}
where $E(z)$ represents the Hubble parameter as a function of redshift. For the SNeIa measurements, the $\chi^2$ function is defined as:
\begin{equation}
    \chi_{\text{SN}}^2(\phi_{\text{s}}^\nu) = \mu_{\text{s}} \mathbf{C}_{\text{s,cov}}^{-1} \mu_{\text{s}}^T,
\end{equation}
where $\mu_{\text{s}}$ is a vector containing the differences between the observed and theoretical distance moduli for each supernova, $\mu_i = \mu_{B,i} - \mathcal{M}$, and $\mathcal{M}$ is a universal free parameter. Note that the covariance matrix $\mathbf{C}_{\text{s,cov}}$ for the SNeIa dataset is not necessarily diagonal. Similarly, A sample of 162 Gamma Ray Bursts (GRB) \cite{GRB} encompassing in the redshift range of $1.44<z<8.1$ has been also assessed. for the GRB measurements, the $\chi^2$ function can be defined as:
\begin{equation}
    \chi_{\text{GRB}}^2(\phi_{\text{g}}^\nu) = \mu_{\text{g}} \mathbf{C}_{\text{g,cov}}^{-1} \mu_{\text{g}}^T,
\end{equation}
Finally, 24 observations of compact radio quasars \cite{quasers} exhibiting in the redshifts of $0.46\leq z\leq 2.76$ where $\mu_{\text{g}}$ is the vector of differences between the observed and theoretical distance moduli for each GRB. Likewise, for the 24 quasar (Q) \cite{quasers} measurement exhibiting in the redshifts of $0.46\leq z\leq 2.76$, the $\chi^2$ function can be defined as:
\begin{equation}
    \chi_{\text{Q}}^2(\phi_{\text{q}}^\nu) = \mu_{\text{q}} \mathbf{C}_{\text{q,cov}}^{-1} \mu_{\text{q}}^T,
\end{equation}
where $\mu_{\text{q}}$ represents the vector of differences between the observed and theoretical distance moduli for each quasar. In the overall analysis.
\subsubsection{Baryon Acoustic Oscillations (BAO)}
\begin{figure}[H]
\centering
\includegraphics[scale=0.63]{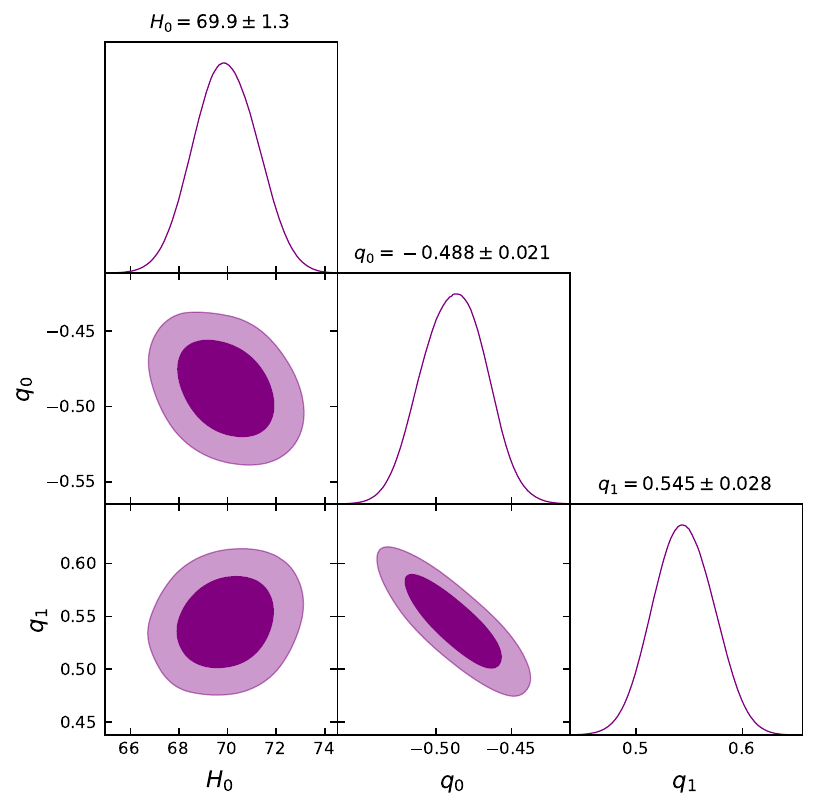}
\caption{This figure corresponds to $ 1\sigma $ and $ 2\sigma $ confidence contours obtained from  $H(z)$ + SNIa + GRB + Q + BAO dataset obtained for the emergent Model 1.}
\label{fig_1}
\end{figure}

\begin{figure}[H]
\centering
\includegraphics[scale=0.63]{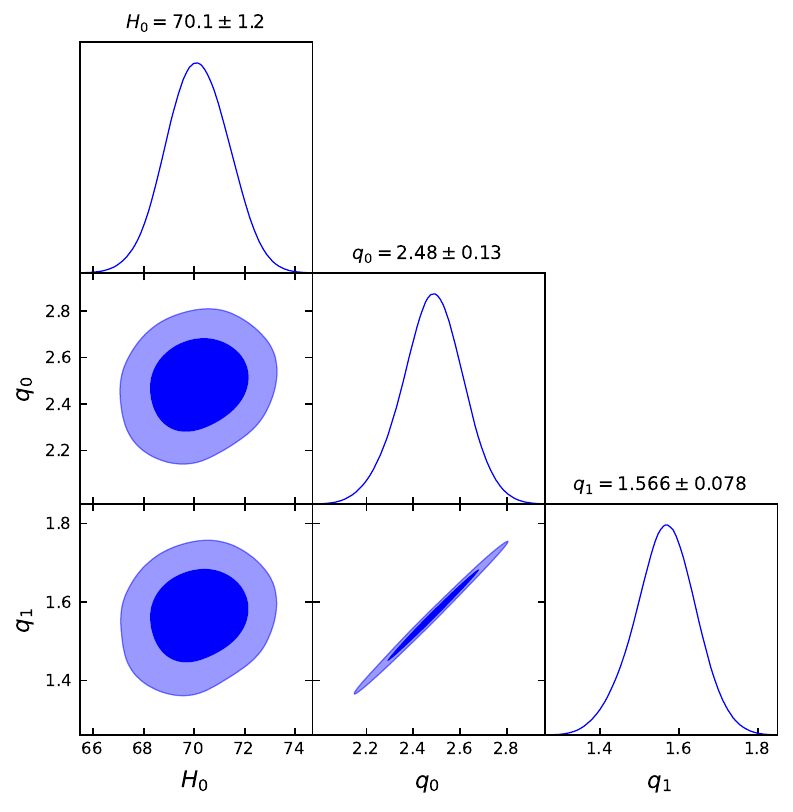}
\caption{This figure corresponds to $ 1\sigma $ and $ 2\sigma $ confidence contours obtained from  $H(z)$ + SNIa + GRB + Q + BAO dataset obtained for the emergent Model 2.}
\label{fig_2}
\end{figure}

\begin{figure}[H]
\centering
\includegraphics[scale=0.63]{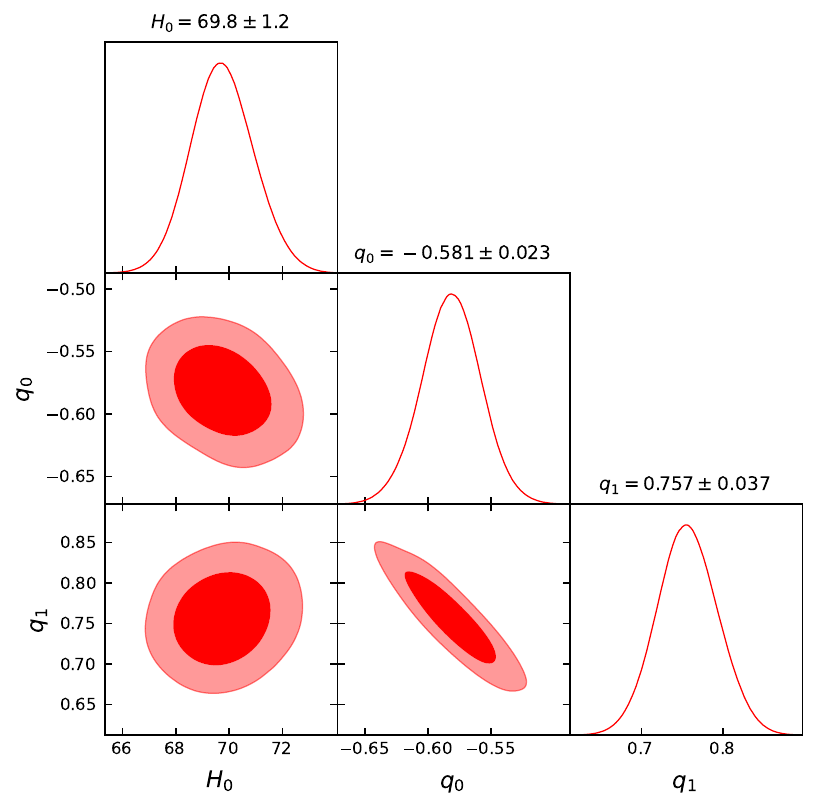}
\caption{This figure corresponds to $ 1\sigma $ and $ 2\sigma $ confidence contours obtained from  $H(z)$ + SNIa + GRB + Q + BAO dataset obtained for the emergent Model 3.}
\label{fig_3}
\end{figure}

The investigation of Baryon Acoustic Oscillations (BAO) involves analyzing a comprehensive dataset consisting of 333 measurements measures from measures from \cite{baonew1,baonew2,bao3,bao4,bao5,bao6,bao7,bao8,bao9,bao10,bao11,bao12}, However, to mitigate potential errors arising from data correlations, we have curated a smaller subset of 17 BAO measurements for our analysis (please see table 1 of this work \cite{benisty2021testing}). This careful selection ensures a more accurate and reliable assessment \cite{Bao1,Bao2}. One of the crucial quantities derived from BAO studies in the transverse direction is represented by $D_H(z)/r_d$, where $D_H(z)$ signifies the comoving angular diameter distance. It can be evaluated using the following equation:
\begin{equation}
D_M = \frac{c}{H_0} S_k\left(\int_0^z \frac{d z^{\prime}}{E\left(z^{\prime}\right)}\right),
\end{equation}
Here, $S_k(x)$ is a function defined as:
\begin{equation}
S_k(x) = \begin{cases}\frac{1}{\sqrt{\Omega_k}} \sinh \left(\sqrt{\Omega_k} x\right) & \text { if } \quad \Omega_k>0 \\ x & \text { if } \quad \Omega_k=0 \\ \frac{1}{\sqrt{-\Omega_k}} \sin \left(\sqrt{-\Omega_k} x\right) & \text { if } \quad \Omega_k<0 .\end{cases}
\end{equation}
Moreover, we consider the angular diameter distance given by $D_A = D_M / (1+z)$, as well as the quantity $D_V(z)/r_d$. The latter is a combination of the BAO peak coordinates and $r_d$, which represents the sound horizon at the drag epoch. Additionally, "line-of-sight" or "radial" observations can be directly obtained from the Hubble parameter using the equation:

\begin{equation}
D_V(z) \equiv \left[z D_H(z) D_M^2(z)\right]^{1/3}.
\end{equation}
By thoroughly analyzing these BAO measurements, we gain valuable insights into the cosmological properties and evolution of the Universe. The careful selection of a smaller dataset allows us to minimize potential errors and consider essential distance measures and observational parameters in our investigation. The total chi-square is therefore given by,
\begin{equation}
\chi^{2}(H_0, q_{0}, q_{1})=\chi_{CC}^{2}+{\chi}_{SNIa}^2 + \chi_{GRB}^{2} + \chi_{Q}^{2} + \chi_{BAO}^{2}.
\end{equation}
The posterior distributions for Models 1, 2, and 3 are illustrated in Figures \ref{fig_1}, \ref{fig_2}, and \ref{fig_3}, respectively, displaying 1$\sigma$ and 2$\sigma$ constraints at 95\% confidence level (CL). The errors on the cosmological parameters for each model are summarized in Table \ref{table1}.\
\begin{table}[]
\begin{center}
\begin{tabular}{|c|c|c|c|}
\hline
\multicolumn{4}{|c|}{MCMC Results} \\ \hline
Model & Parameters & Priors & Bestfit Value \\ \hline
$\Lambda$CDM Model & $H_0$ & [50.,100.] & $69.854848_{\pm 1.259100}^{\pm 1.259100}$ \\ \hline
Model 1 & $H_0$ & [50.,100.] & $69.923922_{\pm 1.283587}^{\pm 2.467758}$ \\ 
& $q_0$ & [-1.,0.] & $-0.487956_{\pm 0.021678}^{\pm 0.038379}$ \\ 
& $q_1$ & [0.,1.] &$0.5445540_{\pm 0.028762}^{\pm 0.0532249}$ \\ \hline
Model 2 & $H_0$ & [50.,100.] & $70.137451_{\pm 1.180809}^{\pm 2.415098}$ \\ 
& $q_0$ & [2.,3.] &$2.482794_{\pm 0.125124}^{\pm 0.273701}$ \\ 
& $q_1$ & [1.,2.] & $1.565706_{\pm 0.076394}^{\pm 0.161332}$ \\ \hline
Model 3 & $H_0$ & [50.,100.] & $69.750246_{\pm 1.129289}^{\pm 2.150918}$ \\ 
& $q_0$ & [-1.,0.] &$-0.581412_{\pm 0.021646}^{\pm 0.046809}$ \\ 
& $q_1$ & [0.,1.] &$0.756614_{\pm 0.036238}^{\pm 0.071081}$ \\ 
\hline
\end{tabular}
\end{center}
\caption{Best fit values of the Model parameters}\label{table1}
\end{table}
\section{Observational, and theoretical comparisons of the Hubble and distance modulus function }\label{sec4}
After determining the free parameters of the respective model, the next step involves comparing the model's predictions with observational data as well as with the standard $\Lambda$CDM model. This comparison allows us to assess the agreement between the model and the data, and evaluate its performance relative to the well-established $\Lambda$CDM framework. By analyzing the consistency or discrepancies between the model's predictions and observational evidence, we can gain insights into the model's viability and its ability to explain the observed phenomena.
\subsection{Comparison with the CC measurements.} 
\begin{figure}[!htb]
   \begin{minipage}{0.49\textwidth}
     \centering
    \includegraphics[scale=0.38]{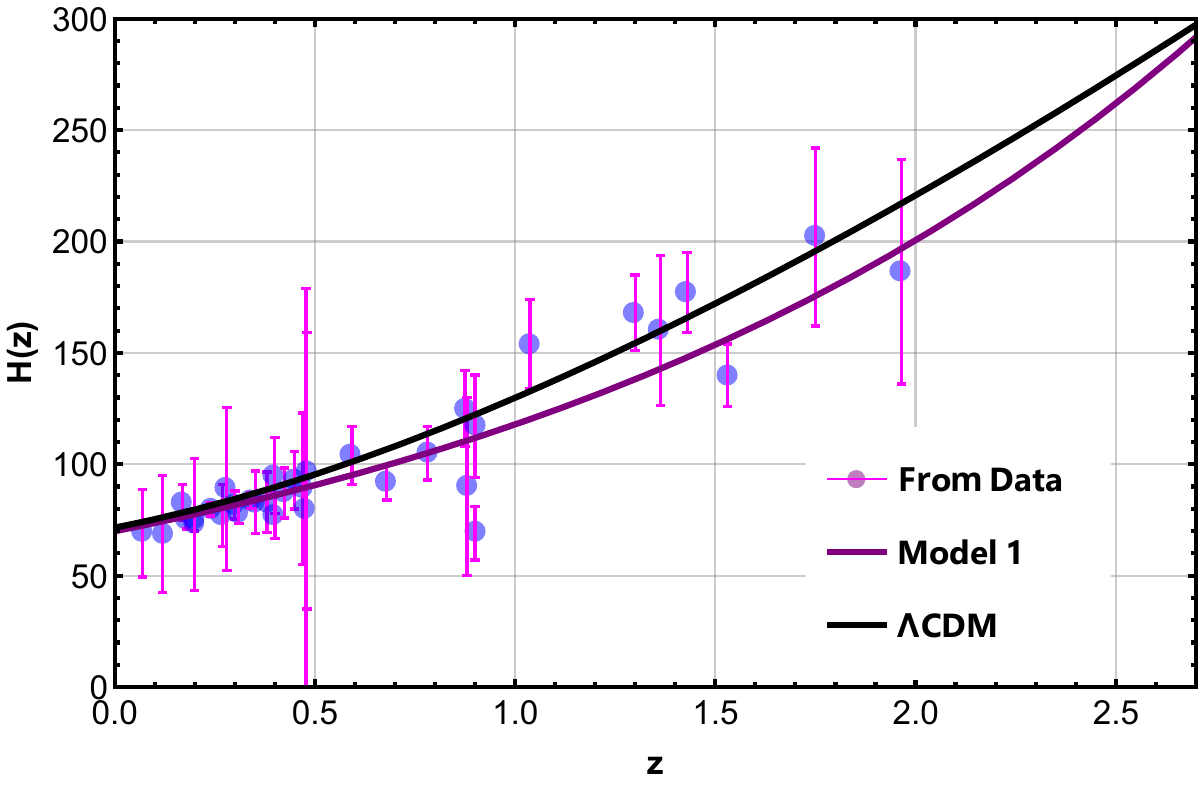}
\caption{The figure shows that the theoretical curve of the Hubble function $H(z)$ of Model 1 is shown in purple line and $\Lambda$CDM Model shown in black line with $\Omega_{\mathrm{m0}}=$ 0.3 and $\Omega_\Lambda =$ 0.7 against 57 $H(z)$ datasets are shown in blue dots with their corresponding error bars shown in purple bars.}\label{H(z)1}
   \end{minipage}\hfill
   \begin{minipage}{0.49\textwidth}
     \centering
   \includegraphics[scale=0.38]{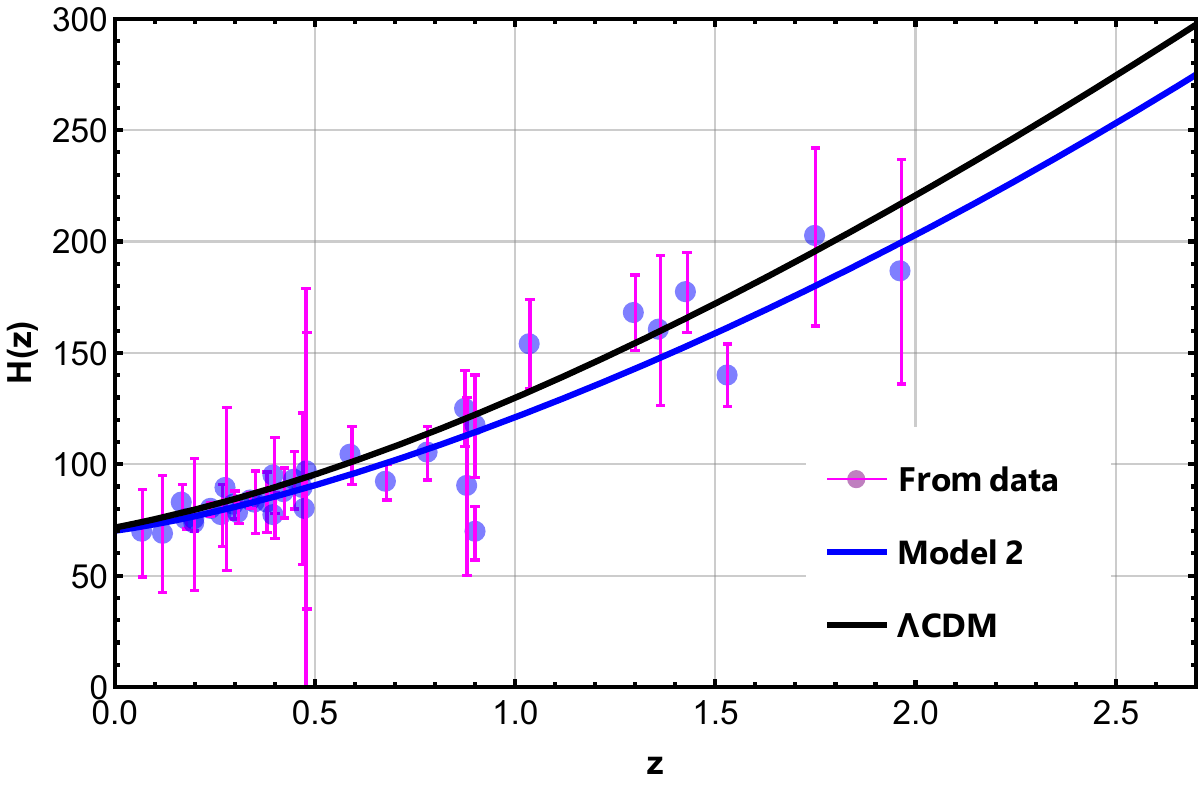}
\caption{The figure shows that the theoretical curve of the Hubble function $H(z)$ of Model 2 is shown in blue line and $\Lambda$CDM Model shown in black line with $\Omega_{\mathrm{m0}}=$ 0.3 and $\Omega_\Lambda =$ 0.7 against 57 $H(z)$ datasets are shown in blue dots with their corresponding error bars shown in purple bars}\label{H(z)2}
   \end{minipage}\hfill
   \begin{minipage}{0.49\textwidth}
     \centering
   \includegraphics[scale=0.38]{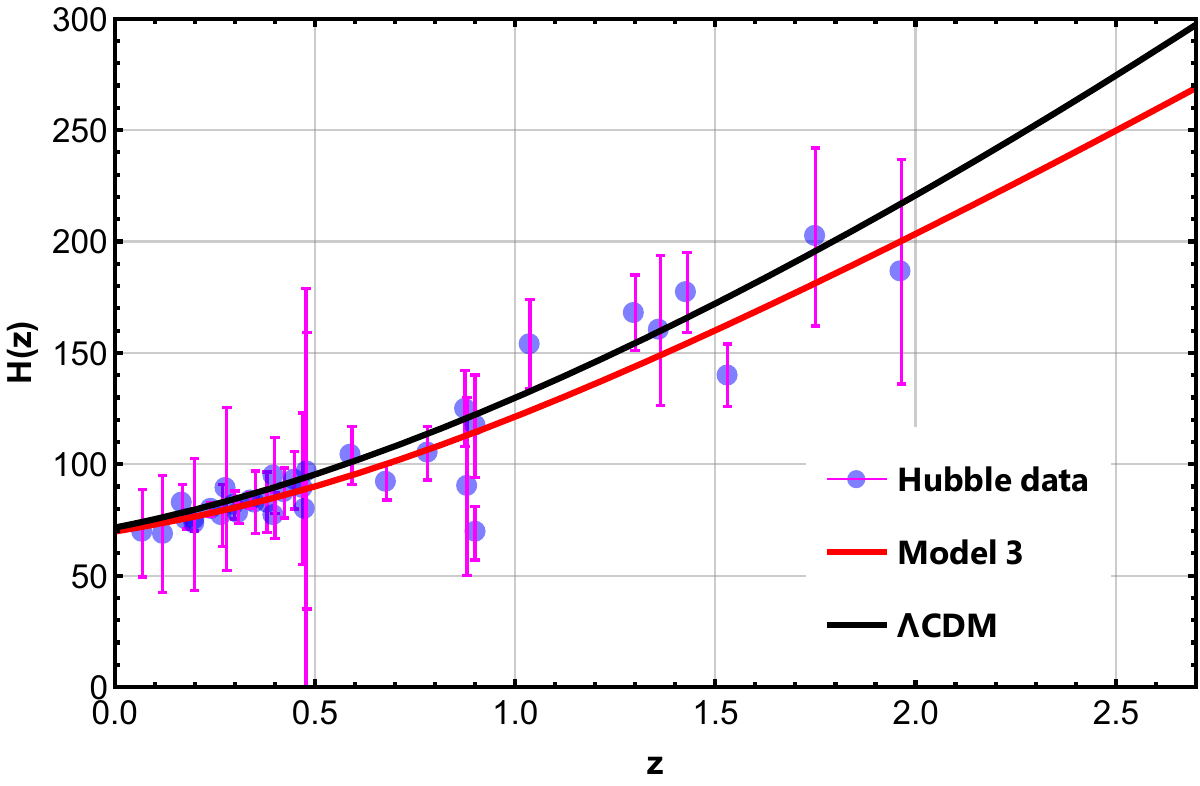}
\caption{The figure shows that the theoretical curve of the Hubble function $H(z)$ of Model 3 is shown in red line and $\Lambda$CDM Model shown in black line with $\Omega_{\mathrm{m0}}=$ 0.3 and $\Omega_\Lambda =$ 0.7 against 57 $H(z)$ datasets are shown in blue dots with their corresponding error bars shown in purple bars}\label{H(z)3}
   \end{minipage}
\end{figure}

In our research article, we conducted a comparison between the Respective Model and 31 measurements of the CC. The Cosmic Chronometers (CC)  measurements are represented by blue dots, accompanied by their corresponding error bars depicted in purple bars. Additionally, we included the $\Lambda$CDM model for reference. The results of this comparison are illustrated in Figures.~\ref{H(z)1}, \ref{H(z)2} and \ref{H(z)3}. Notably, our analysis reveals that the Respective Model aligns well with the CC measurements, indicating a good fit between the model and the observed data.
\subsection{Comparison with the type Ia supernova dataset.} 
\begin{figure}[H]
\centering
\includegraphics[scale=0.38]{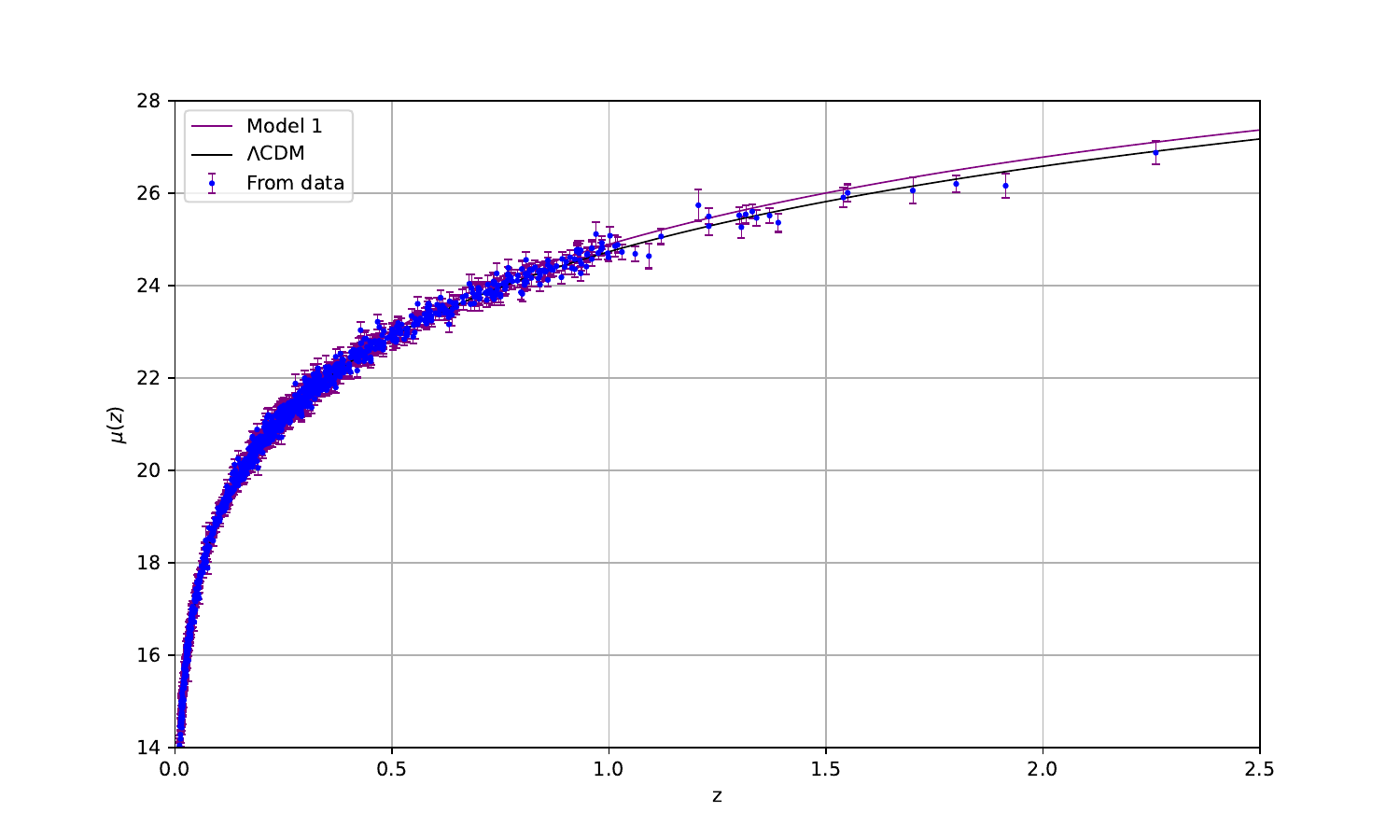}
\caption{Theoretical curve of distance modulus $\protect\mu(z) $ of the Model 1 is shown in purple line and the $\Lambda$CDM Model is shown in the black line with $\Omega_{\mathrm{m0}}=$ 0.3 and $\Omega_\Lambda =$ 0.7 against type Ia supernova data are shown in blue dots with their corresponding errors bars n purple bars}
\label{mu(z)1}
\end{figure}

\begin{figure}[H]
\centering
\includegraphics[scale=0.38]{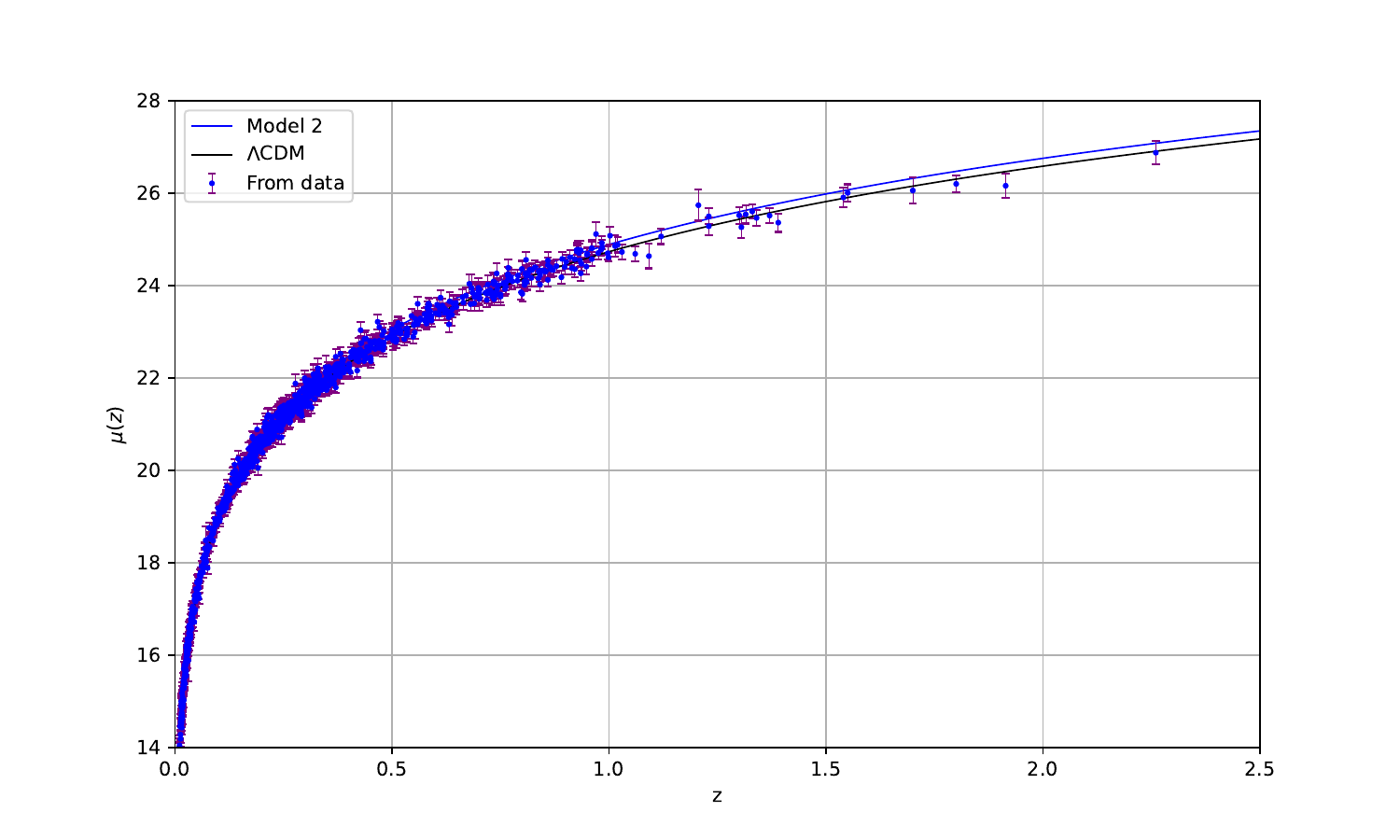}
\caption{Theoretical curve of distance modulus $\protect\mu(z) $ of the Model 2 is shown in blue line and the $\Lambda$CDM Model is shown in the black line with $\Omega_{\mathrm{m0}}=$ 0.3 and $\Omega_\Lambda =$ 0.7 against type Ia supernova data are shown in blue dots with their corresponding errors bars n purple bars}
\label{mu(z)2}
\end{figure}

\begin{figure}[H]
\centering
\includegraphics[scale=0.38]{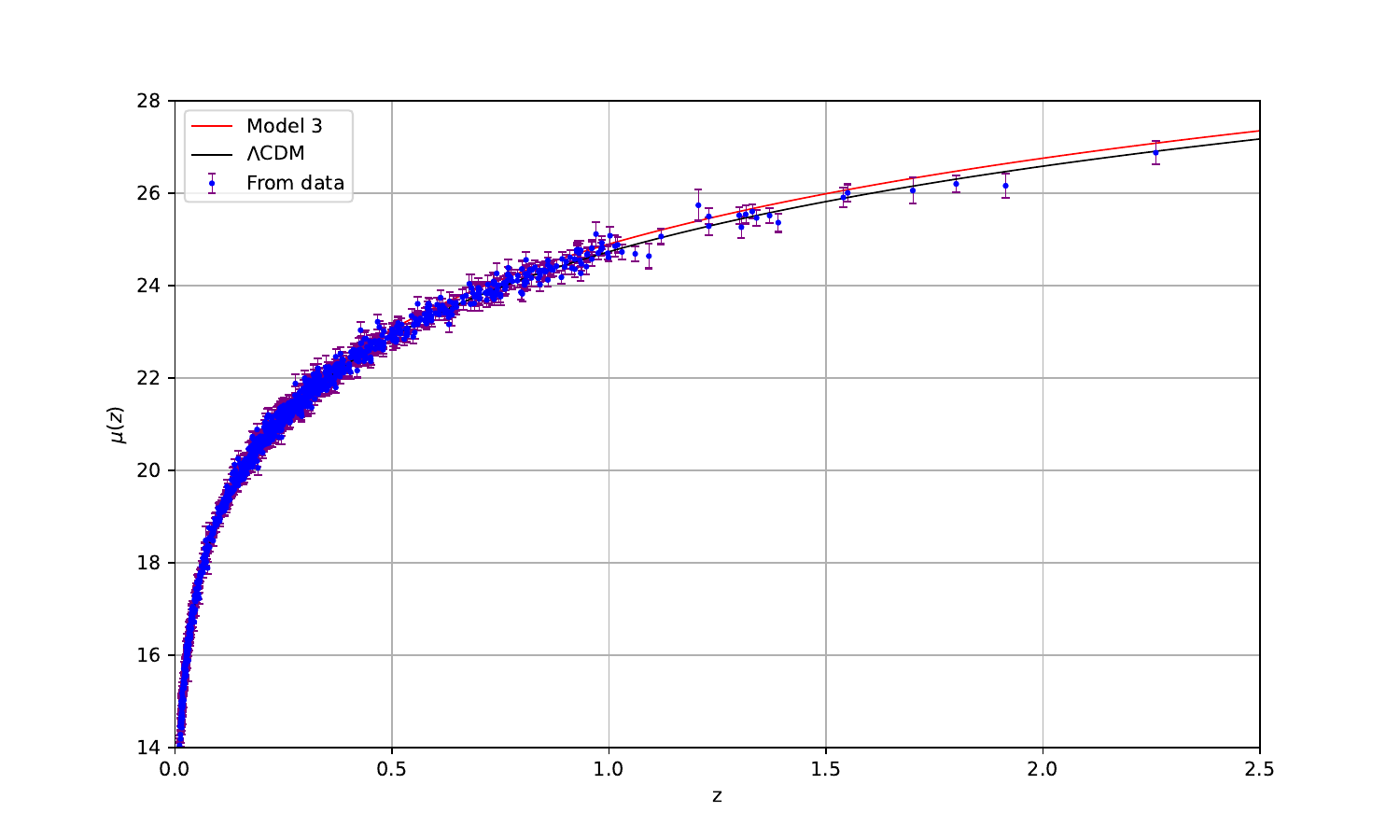}
\caption{Theoretical curve of distance modulus $\protect\mu(z) $ of the Model 3 is shown in red line and the $\Lambda$CDM Model is shown in the black line with $\Omega_{\mathrm{m0}}=$ 0.3 and $\Omega_\Lambda =$ 0.7 against type Ia supernova data are shown in blue dots with their corresponding errors bars n purple bars}
\label{mu(z)3}
\end{figure}

We compare the distance modulus $\mu(z)$ of the Respective Model with the Pantheon data, considering 1$\sigma$ and 2$\sigma$ error bands. By analyzing Figures~ \ref{mu(z)1}, \ref{mu(z)2}, and \ref{mu(z)3}, it becomes evident that the Respective Model is in excellent agreement with the Pantheon distance modulus. The Respective Model, which incorporates 1048 observation points, provides a remarkably accurate fit to the Pantheon data. The model's predictions closely match the observed values, suggesting its efficacy in describing the cosmological phenomena related to the distance modulus. The agreement between the Respective Model and the Pantheon data highlights the model's reliability and its potential to contribute to new research endeavors in the field of cosmology.

\section{Cosmographic Parameters}\label{sec5}
Cosmographic parameters play a fundamental role in our understanding of the Universe's expansion and its underlying dynamics \cite{visser2004jerk,visser2004jerk3,visser2005cosmography,visser2010cosmographic,lobo2020cosmographic}. These parameters provide insights into the evolution of the cosmic scale factor, as well as the Hubble parameter and its derivatives. By analyzing the cosmographic parameters, such as the deceleration parameter, jerk parameter, and snap parameter, we can probe the nature of dark energy, test different cosmological models, and explore the implications for the Universe's past and future behavior. The study of cosmographic parameters opens up new avenues for investigating the fundamental properties of our cosmos and deepening our understanding of its evolution.
\subsection{Decceleration Parameter}
In the field of cosmology, the deceleration parameter, denoted as $q$, plays a crucial role in understanding the dynamics of the expanding Universe. It is defined as the negative ratio of the second time derivative of the scale factor, representing the rate of change of the Universe's expansion, to the square of the Hubble parameter multiplied by the scale factor itself. In mathematical terms, the deceleration parameter is given by $q = -\frac{\ddot{a}a}{\dot{a}^2}$, where $a(t)$ represents the scale factor of the Universe as a function of time. The deceleration parameter's sign provides valuable insights into the nature of cosmic expansion. A positive deceleration parameter ($q > 0$) indicates that the Universe is decelerating, while a negative value ($q < 0$) signifies an accelerating expansion. The deceleration parameter thus serves as a fundamental quantity for discerning the overall dynamics and behavior of the evolving cosmos.
\begin{figure}[H]
\centering
\includegraphics[scale=0.42]{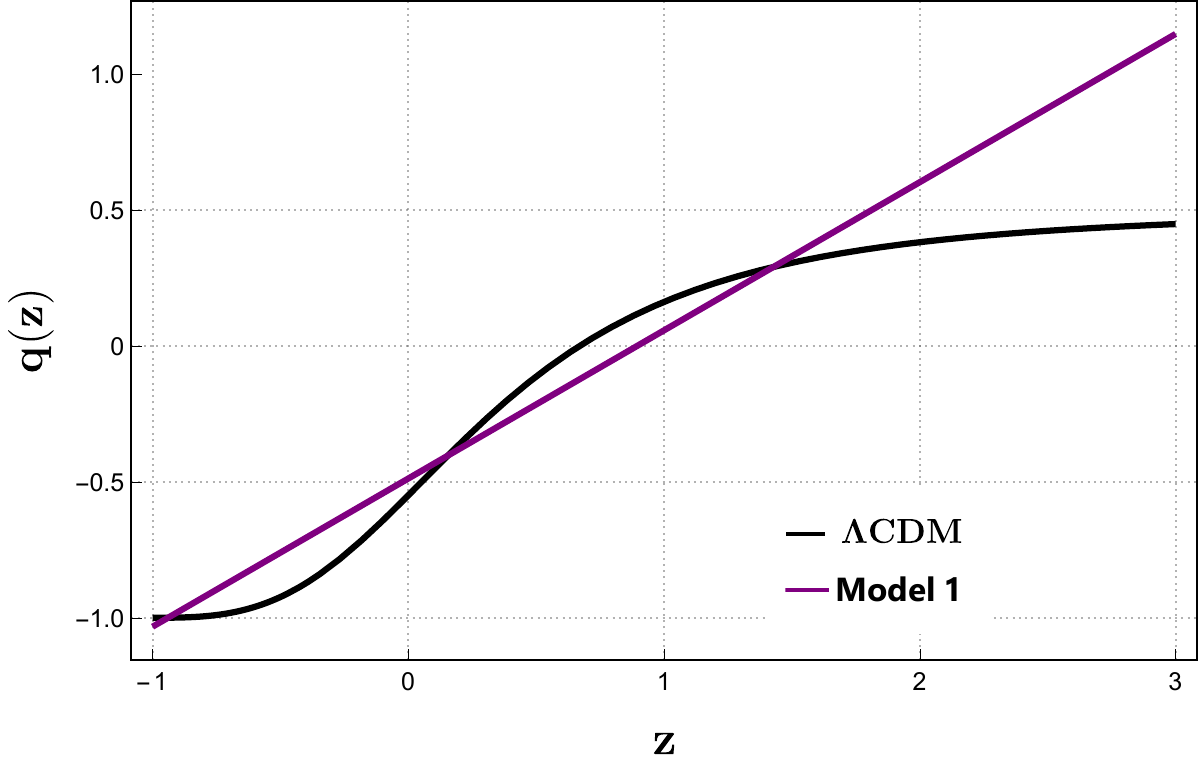}
\caption{This figure provides visual perception between respective and $\Lambda$CDM Model of deceleration Parameter.}
\label{qz1}
\end{figure}

\begin{figure}[H]
\centering
\includegraphics[scale=0.42]{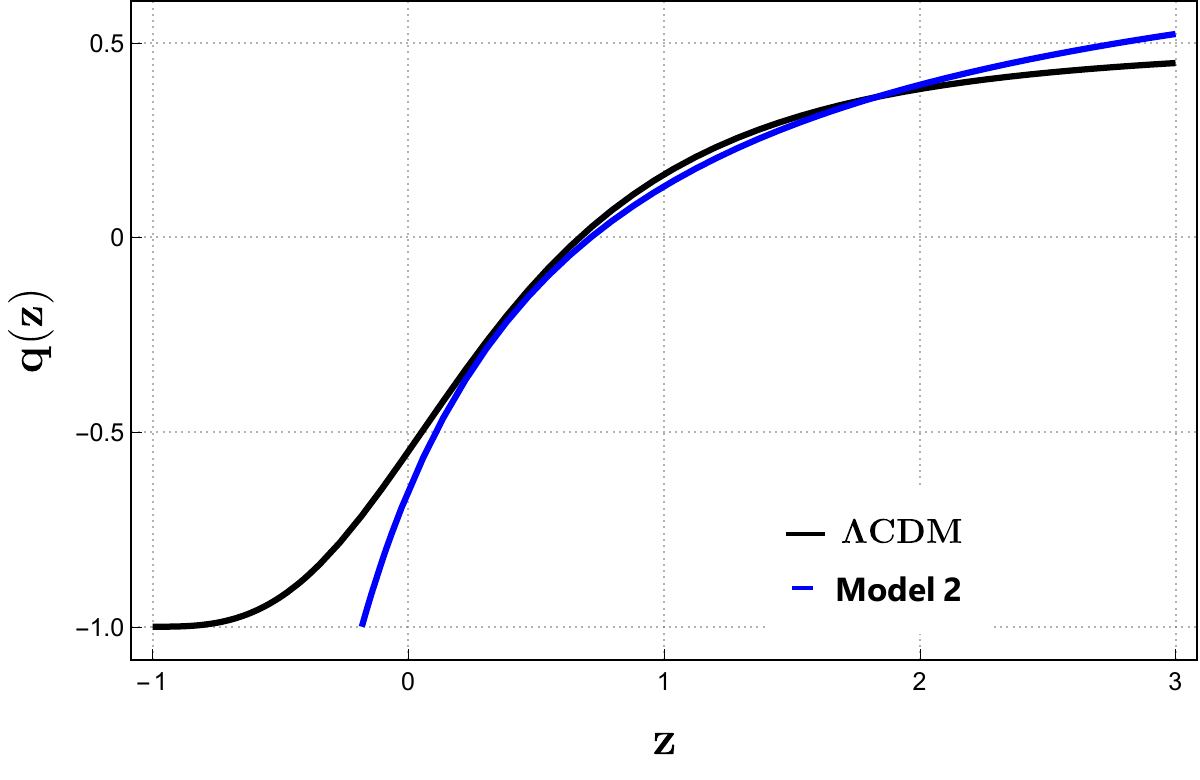}
\caption{This figure provides visual perception between respective and $\Lambda$CDM Model of deceleration Parameter.}
\label{qz2}
\end{figure}

\begin{figure}[H]
\centering
\includegraphics[scale=0.42]{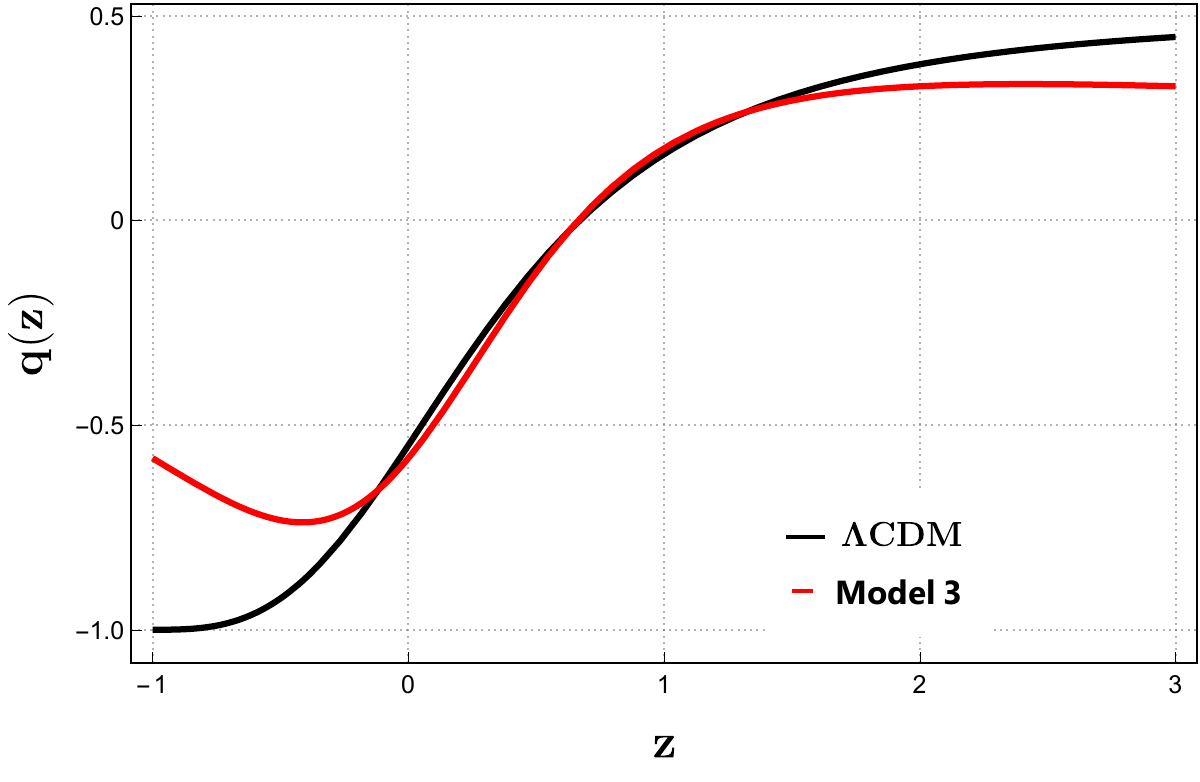}
\caption{This figure provides visual perception between respective and $\Lambda$CDM Model of deceleration Parameter.}
\label{qz3}
\end{figure}

\subsection{Jerk Parameter}
The deceleration parameter plays a crucial role in characterizing the dynamics of the Universe's expansion. In terms of the deceleration parameter $q$, the jerk parameter $j$ provides additional insights into the acceleration behavior. It is defined as the ratio of the third derivative of the scale factor with respect to cosmic time to the cube of the first derivative of the scale factor. For a given cosmological model, the jerk parameter can be expressed in terms of the deceleration parameter as $j = [(1+z) \frac{dq}{dz} + q(2q+1)]$. In the case of the $\Lambda$CDM model, which assumes a cosmological constant, the deceleration parameter remains constant and equal to $q = -1$, resulting in $j = 1$. This indicates that the $\Lambda$CDM model exhibits a constant rate of acceleration. Understanding the behavior of the deceleration parameter and its relationship to the jerk parameter provides valuable insights into the dynamics of the Universe's expansion and the nature of dark energy.

\begin{figure}[H]
\centering
\includegraphics[scale=0.42]{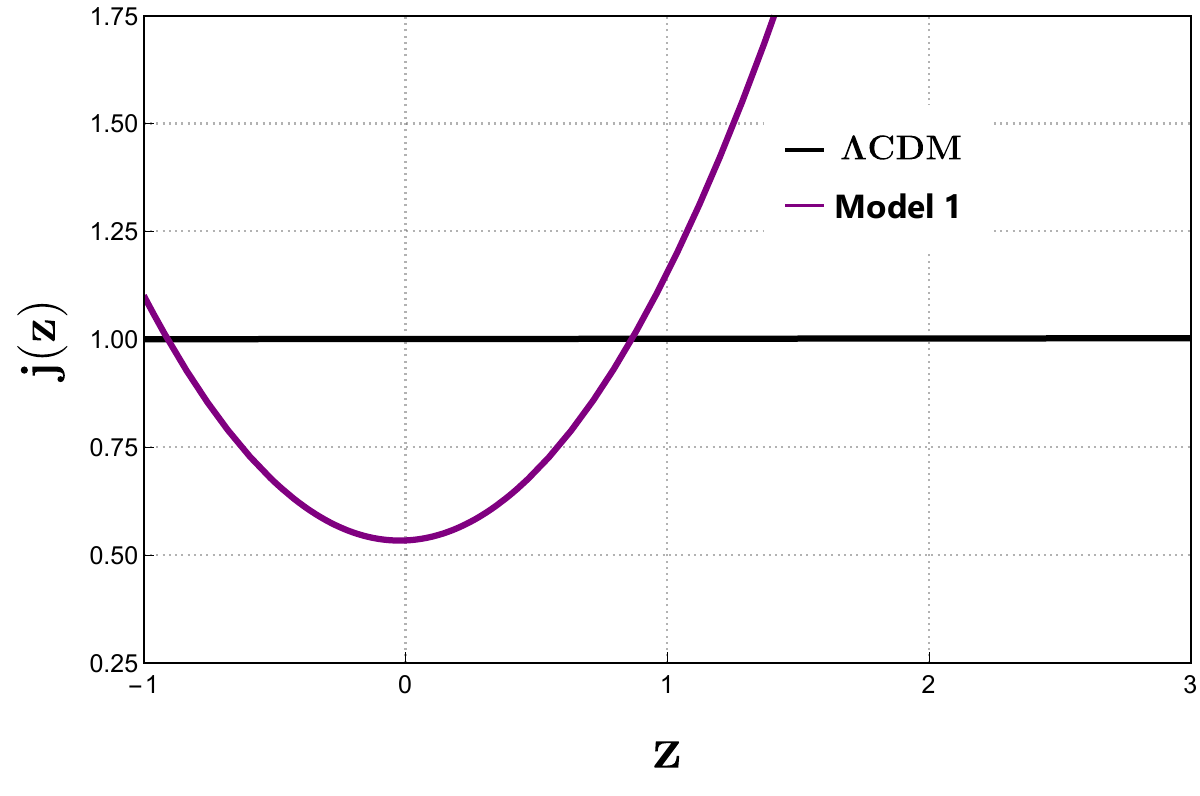}
\caption{This figure provides visual perception between respective and $\Lambda$CDM Model of jerk Parameter.}
\label{jz1}
\end{figure}

\begin{figure}[H]
\centering
\includegraphics[scale=0.42]{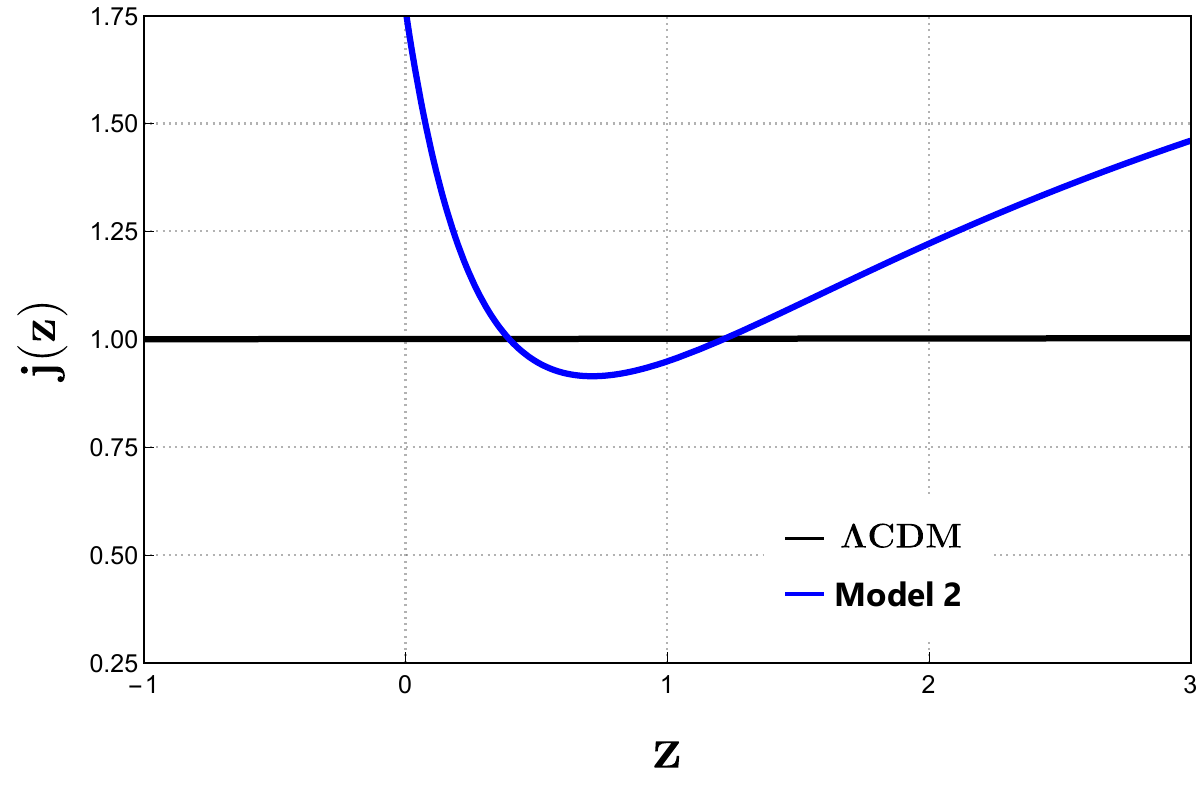}
\caption{This figure provides visual perception between respective and $\Lambda$CDM Model of jerk Parameter.}
\label{jz2}
\end{figure}

\begin{figure}[H]
\centering
\includegraphics[scale=0.42]{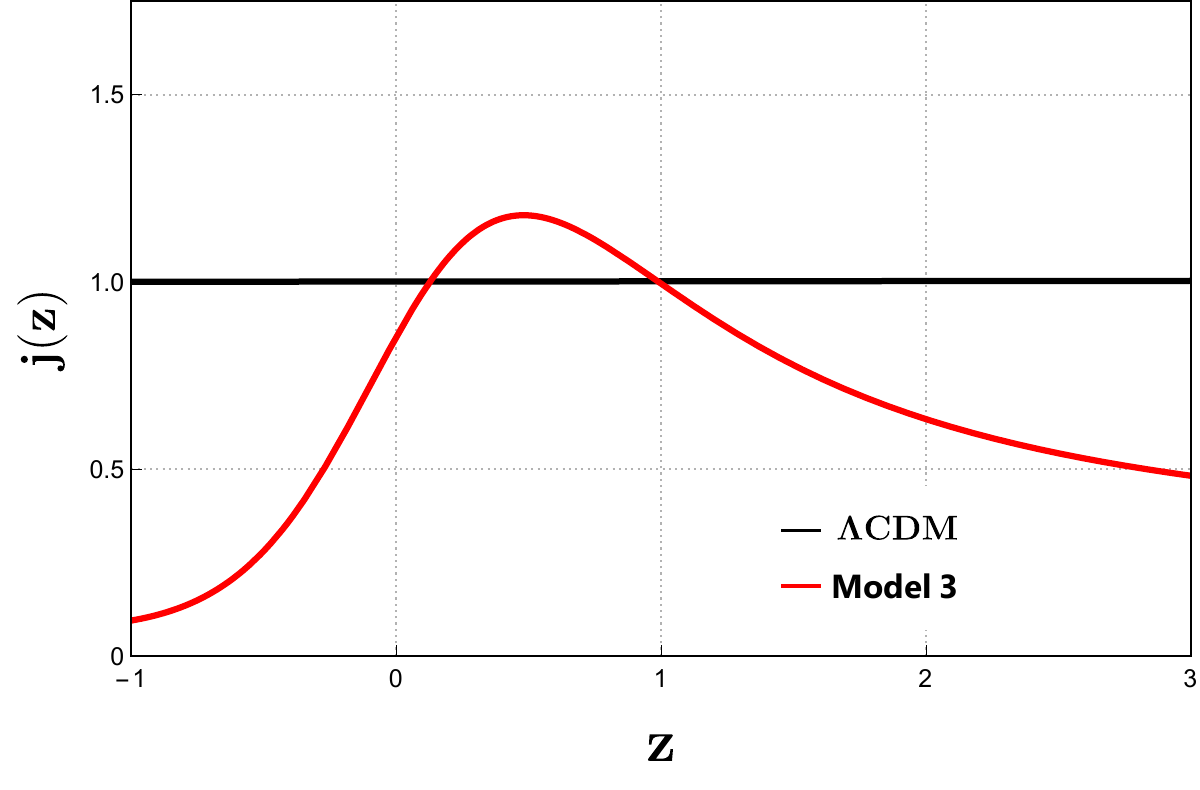}
\caption{This figure provides visual perception between respective and $\Lambda$CDM Model of jerk Parameter.}
\label{jz3}
\end{figure}

\section{Statefinder Diagnostics}\label{sec6}
The two most important geometric parameters are the Hubble constant $H_0$ and the current value of the deceleration parameter $q_0$. Because accurate measurements of the Universe's expansion law in the past are now possible (e.g., using the luminosity distance to distant supernovae), these variables should be generalized to the Hubble parameter $H(z)$ and the deceleration parameter $q(z)$. However, the need to consider more general models of dark energy than a cosmological constant and the remarkable increase in the accuracy of cosmological observational data over the last few years compels us to go beyond these two important quantities. For this reason, the authors in \cite{29,30} proposed a new geometrical diagnostic pair for dark energy. For separating a wide range of dark energy models from existing models like $\Lambda$CDM, SCDM, holographic dark energy, Chaplygin Gas, and Quintessence, statefinder diagnostics offer a potent geometrical diagnostic method. The approach yields two parametric solutions, one between $r$ (or $j$) and $s$ and the other between $r$ and $q$, where these parameters are constructed from higher order pair of the scale factor and are defined as follows:
\begin{eqnarray}
	r = \frac{\dddot a}{a H^{3}}    
\end{eqnarray}
\begin{eqnarray}
	s = \frac{r-1}{3(q-\frac{1}{2})}
\end{eqnarray}
From the expression above, we have a relation for $r$ and $q$:
\begin{eqnarray}
	r = q - 2q^{2} - \frac{\dot q}{H}
\end{eqnarray}
The diagnostic pairs $\{s,r\}$ and $\{q,r\}$ are known as the Statefinder pair. The statefinder diagnostics pairs with different trajectories in the $s-r$ and $q-r$ planes are plotted to see the temporal evolution of various dark energy models. Different points in this parametric solution correspond to different dark energy models, such as $\Lambda$CDM relates to $(s = 0;\  r = 1)$, holographic dark energy relates to $(s = 2/3;\  r = 1)$, Chaplygin Gas relates to $(s < 0;\  r > 1)$, SCDM relates to $(s = 1;\  r = 1)$, Quintessence relates to $(s > 0;\  r < 1)$, and so on. The fixed points in this context are generally considered to be $\{s,r\}$ = (0,1) and $\{s,r\}$ = (1,1) in FLRW background, and deviations from these fixed points of any dark energy model are analyzed.
\begin{figure}[H]
\centering
\includegraphics[scale=0.44]{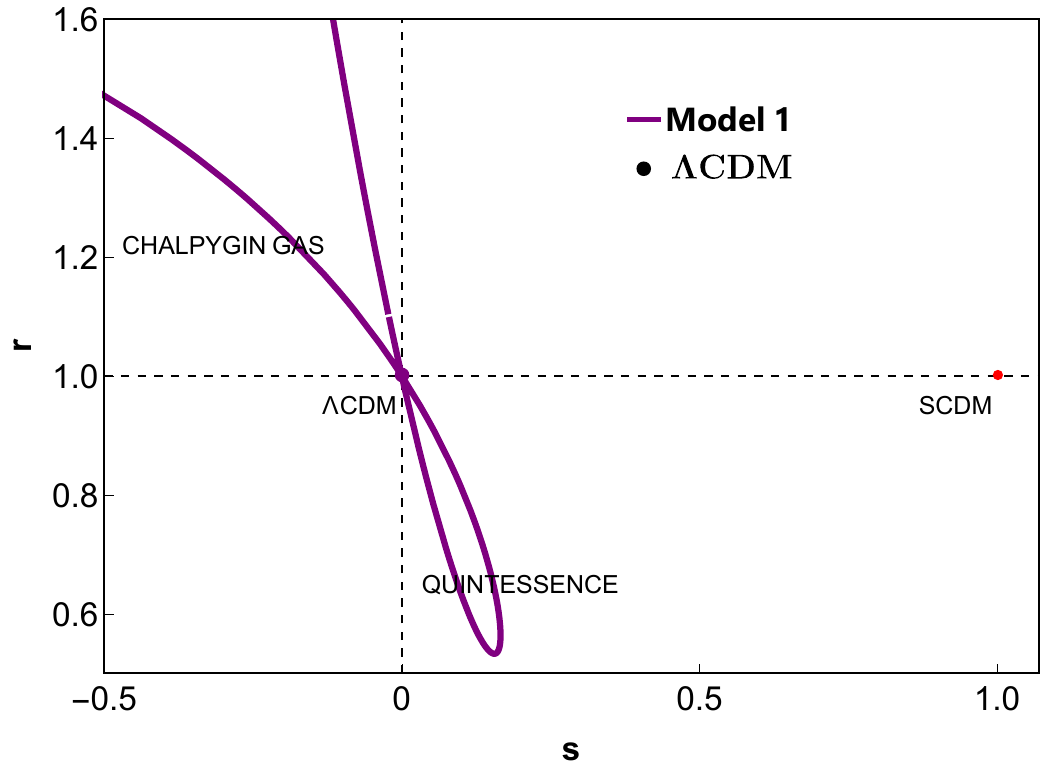}
\caption{Behavior of $\left\{r, s\right\}$ plane Model 1}
\label{rs}
\end{figure}
\begin{figure}[H]
\centering
\includegraphics[scale=0.44]{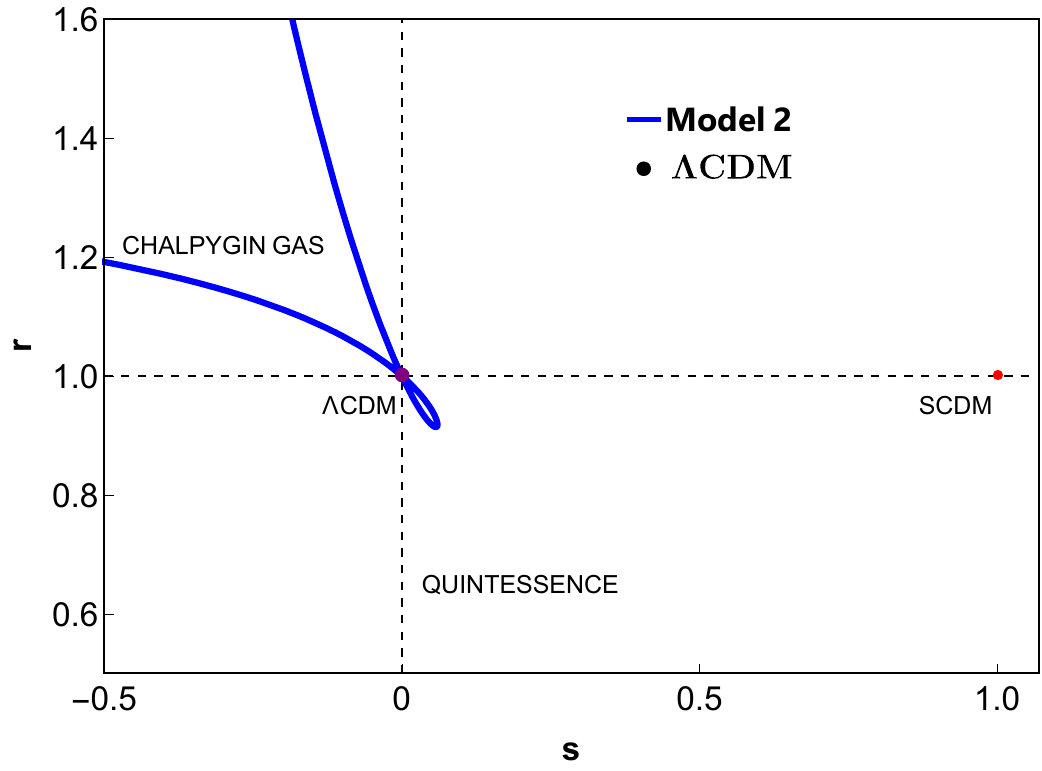}
\caption{Behavior of $\left\{r, s\right\}$ plane Model 2}
\label{rs2}
\end{figure}
\begin{figure}[H]
\centering
\includegraphics[scale=0.44]{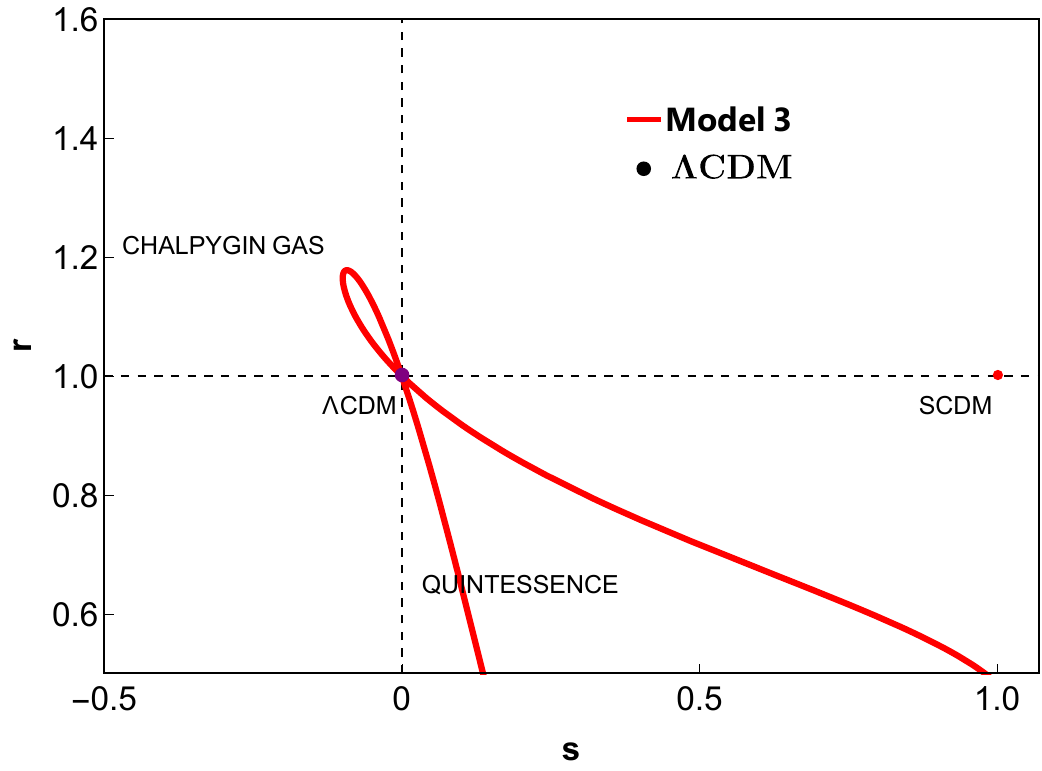}
\caption{Behavior of $\left\{r, s\right\}$ plane Model 3}
\label{rs3}
\end{figure}
\begin{figure}[H]
\centering
\includegraphics[scale=0.44]{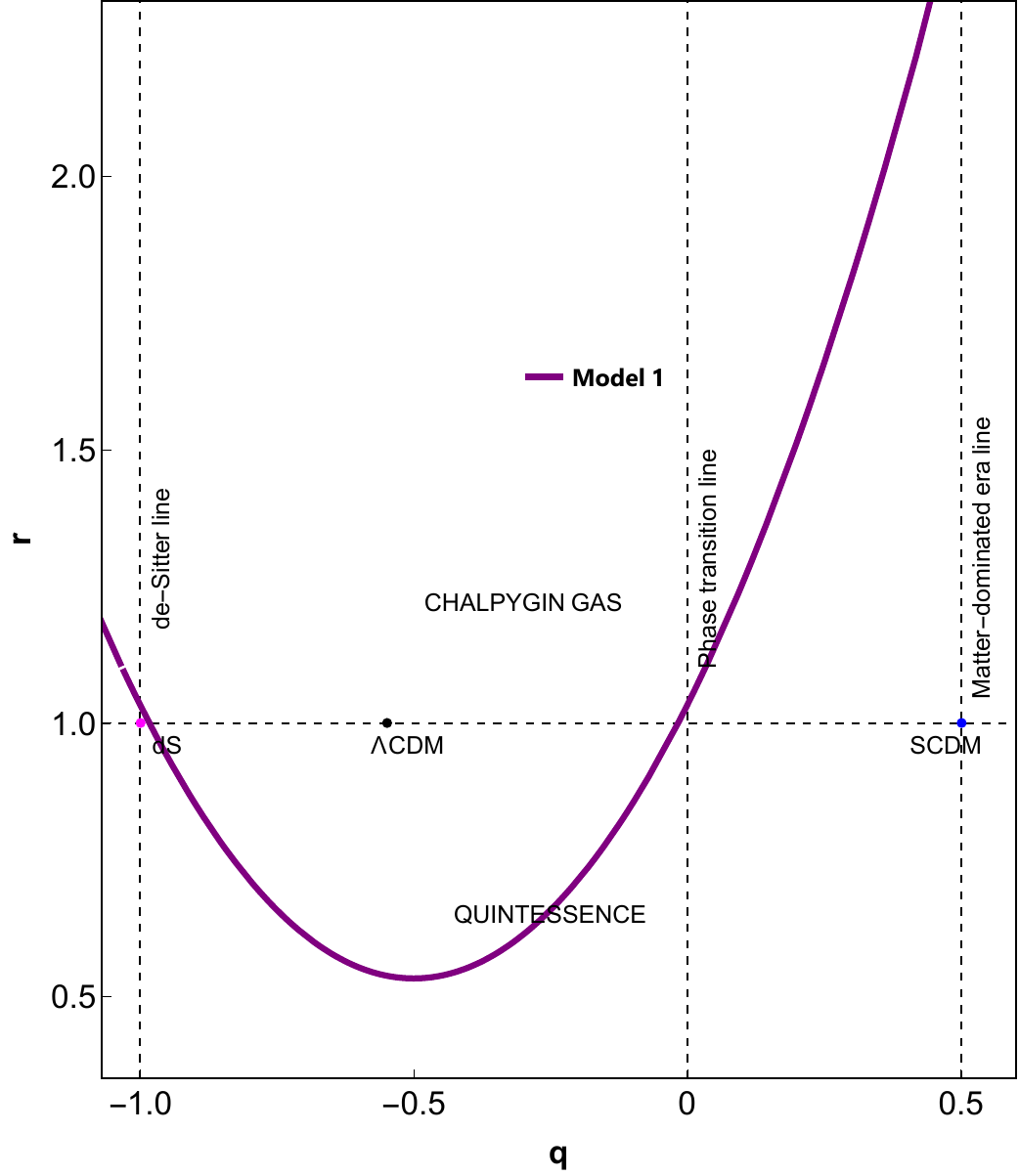}
\caption{Behavior of $\left\{r, q\right\}$ profile Model 1}
\label{rq}
\end{figure}
\begin{figure}[H]
\centering
\includegraphics[scale=0.44]{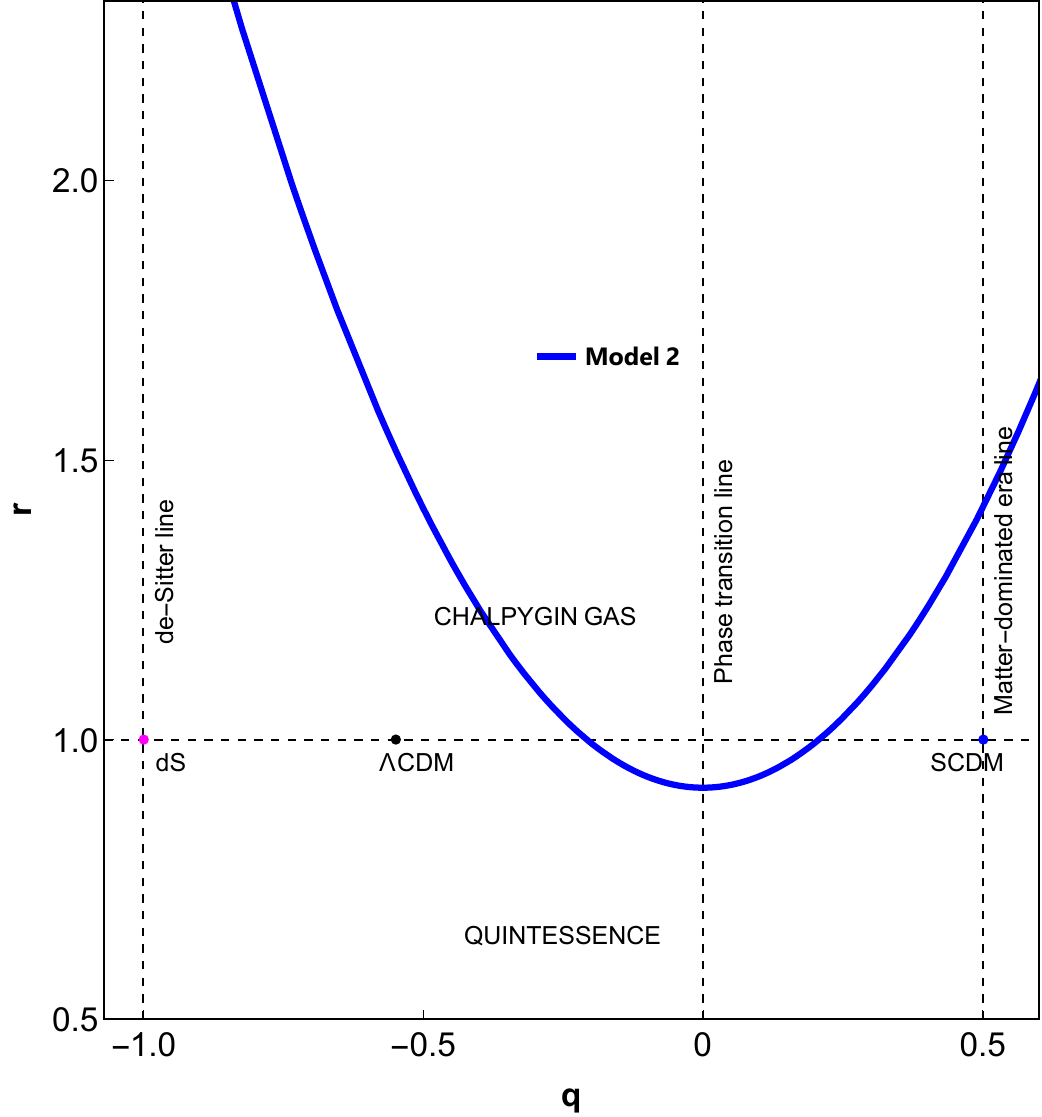}
\caption{Behavior of $\left\{r, q\right\}$ profile Model 2}
\label{rq2}
\end{figure}
\begin{figure}[H]
\centering
\includegraphics[scale=0.44]{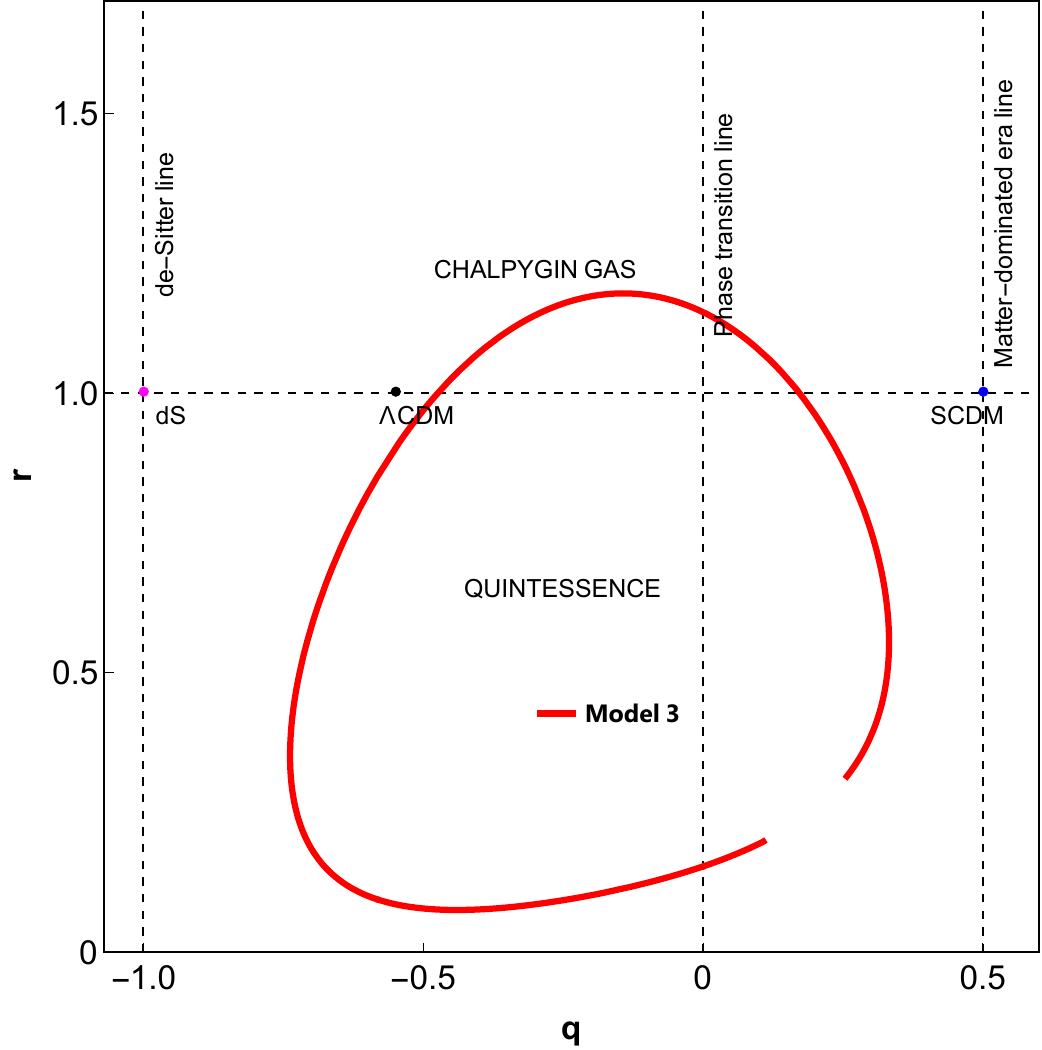}
\caption{Behavior of $\left\{r, q\right\}$ profile Model 3}
\label{rq3}
\end{figure}

\section{$Om(z)$ Diagnostics}\label{sec7}
Om diagnostic is another tool introduced in \cite{31,32} uses the Hubble parameter to provide a null test of the $\Lambda$CDM model. Om diagnostic, like statefinder diagnostic, is an effective method for distinguishing various DE models from CDM models based on slope variation. The positive diagnostic slope indicates a Quintessence nature ($\omega\ge$ -1), the negative diagnostic slope indicates a Phantom nature ( $\omega\le$ -1), and the constant slope with respect to redshift suggests that the nature of dark energy coincides with that of the cosmological constant ( $\omega$= -1). The $Om(z)$ for a flat Universe is defined as:
\begin{eqnarray}
	Om(z) = \frac{(\frac{H_z}{H_0})^{2} -1}{(1+z)^{3}-1}
\end{eqnarray}
The plots for the cosmic jerk, statefinder and om diagnostics for all the models are shown below : The transition of all the models from various types of dark energies can be studied by looking at the $s-r$ and $q-r$ plots. Model 1 begins in the Chaplygin gas region where $r>1$ and $s<0$ and makes a transition to the Quintessence region, where $r<1$ and $s>0$ by crossing the intermediate $\Lambda$CDM point $(1,0)$ during evolution. Model 2 stays in the Quintessence region for all the values of $s$ and $r$. Model 3 shows the opposite behavior to model 1, where the graph begins in the Quintessence region and makes a transition to the Chaplygin gas region after crossing the $\Lambda$CDM point. The temporal evolution of models can be studied by looking at the $q-r$ plots, where the dashed line describes the behavior of $\Lambda$CDM model below, which lies the Quintessence region and, on top of which lies the Chaplygin gas region.
\begin{figure}[H]
\centering
\includegraphics[scale=0.42]{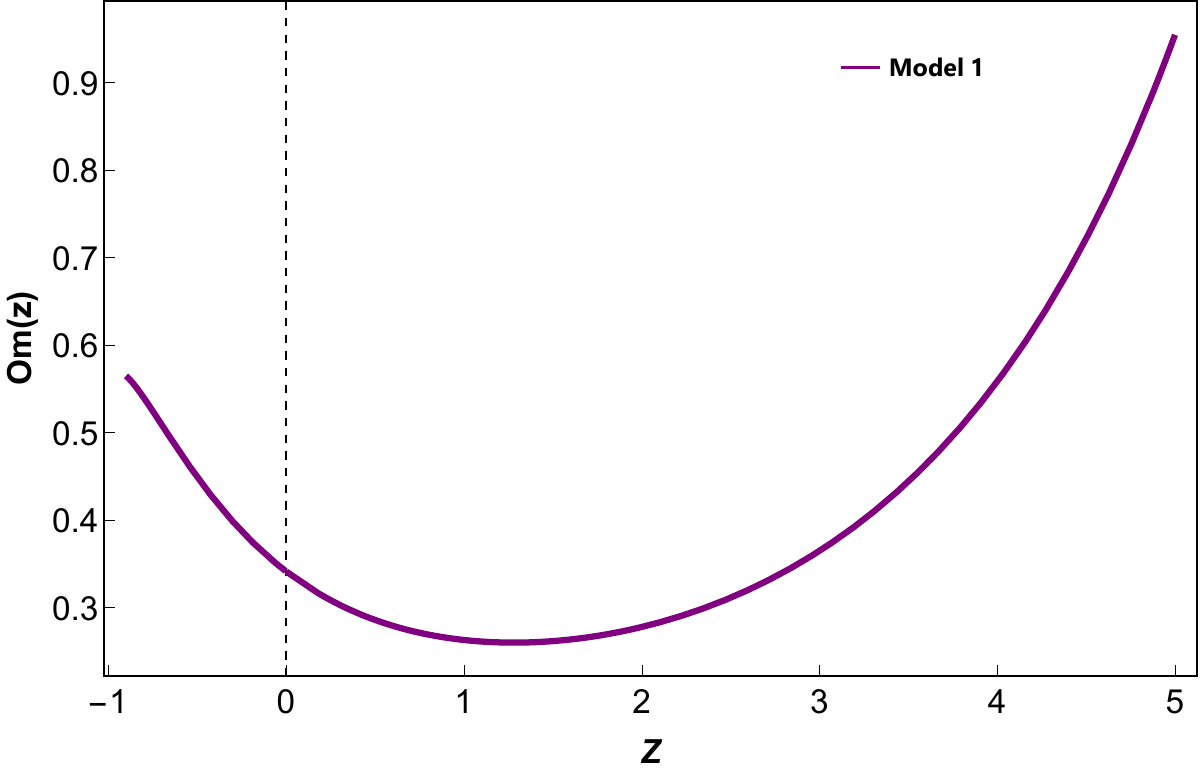}
\caption{This figure shows the $Om(z)$ with respect to redshift.}
\label{Om1}
\end{figure} 

\begin{figure}[H]
\centering
\includegraphics[scale=0.42]{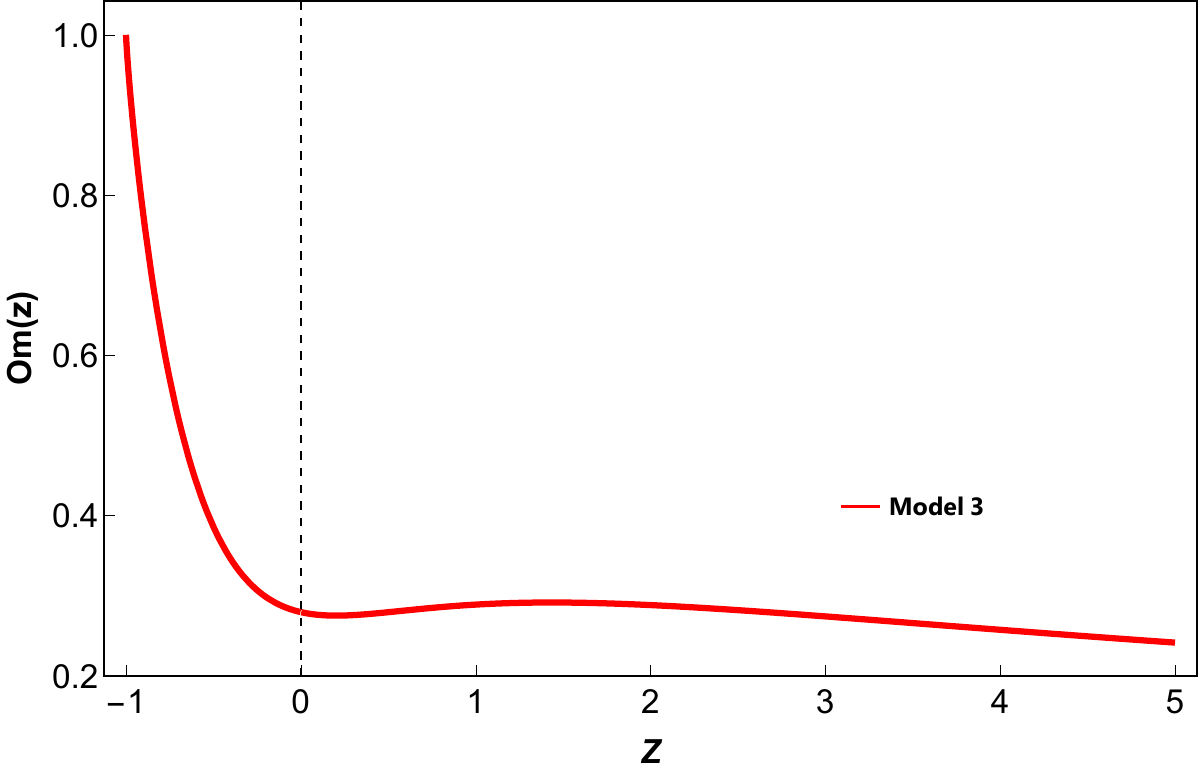}
\caption{This figure shows the $Om(z)$ with respect to redshift.}
\label{Om2}
\end{figure} 

\begin{figure}[H]
\centering
\includegraphics[scale=0.42]{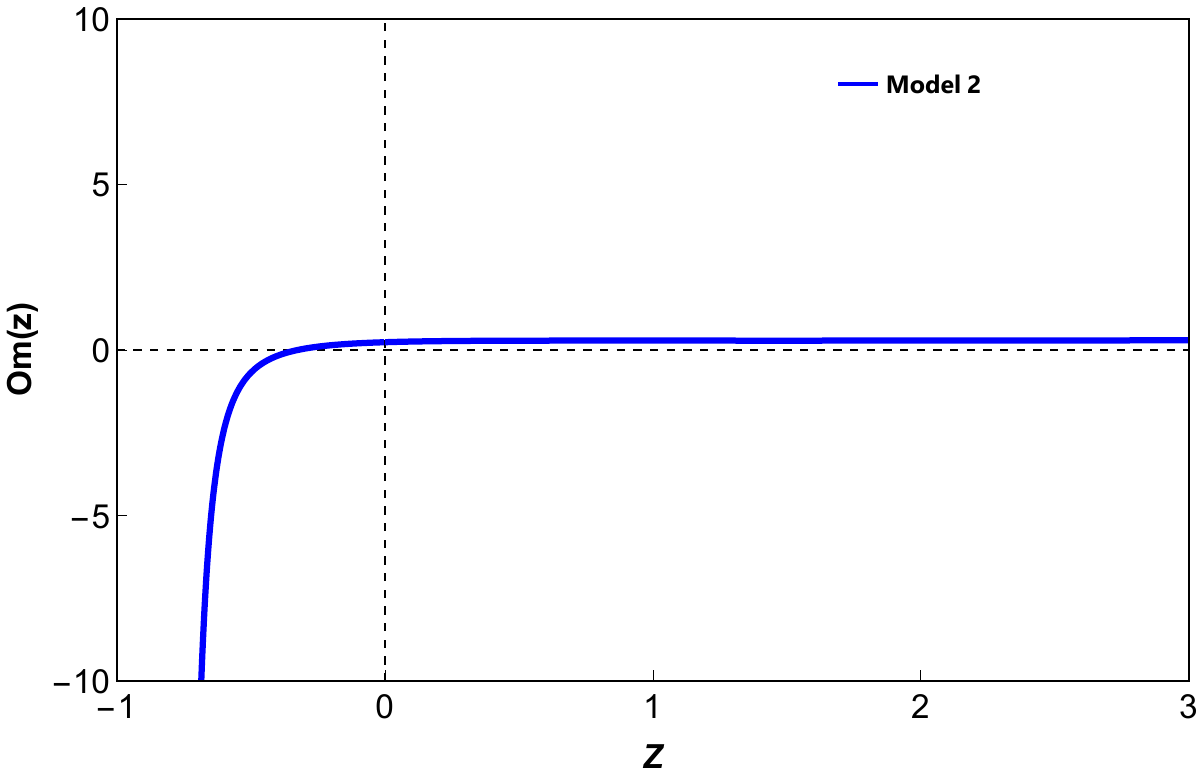}
\caption{This figure shows the $Om(z)$ with respect to redshift.}
\label{Om3}
\end{figure} 

\section{Information Criteria}\label{sec8}
The chi-square statistic is commonly used to measure the goodness of fit between a model and observed data. The minimum chi-square, denoted as ${{\chi}^2_{\text{min}}}$, represents the smallest value obtained for the chi-square statistic in a given analysis. It is calculated as the sum of the squared differences between the observed data and the model's predictions, divided by the uncertainties associated with the data. The formula for ${{\chi}^2_{\text{min}}}$ is
\begin{equation}
{{\chi}^2_{\text{min}}} = \sum \frac{(O_i - E_i)^2}{\sigma_i^2},
\end{equation}
where $O_i$ represents the observed data, $E_i$ denotes the model's predicted values, and $\sigma_i$ represents the uncertainties associated with the data. The reduced chi-square is a normalized version of the chi-square statistic that accounts for the degrees of freedom in the analysis \cite{tauscher2018new}. It is obtained by dividing ${{\chi}^2_{\text{min}}}$ by the number of degrees of freedom, typically denoted as $\nu$. The formula for ${{\chi}^2_{\text{red}}}$ is:
\begin{equation}
{{\chi}^2_{\text{red}}} = \frac{{{\chi}^2_{\text{min}}}}{\nu},
\end{equation}
where $\nu$ is the difference between the number of data points and the number of free parameters in the model. The Akaike Information Criteria (AIC) is an information-theoretic criterion used to compare the relative quality of different models \cite{vrieze2012model,tan2012reliability,rezaei2021comparison}. It takes into account both the goodness of fit and the complexity of the model. AIC is calculated using the formula:
\begin{equation}
\text{AIC} = -2 \ln({\cal L}_{\text{max}}) + 2 \kappa,
\end{equation}
where ${\cal L}_{\text{max}}$ represents the maximum likelihood function and $\kappa$ is the number of parameters in the model. The difference in AIC values, denoted as $\Delta\text{AIC}$, is used to compare the relative support of different models \cite{arevalo2017aic}. It is calculated as the difference between the AIC value of a specific model and the minimum AIC value among all the models considered. The formula for $\Delta\text{AIC}$ is:
\begin{equation}
\Delta\text{AIC} = \text{AIC}_{\text{model}} - \text{AIC}_{\text{min}},
\end{equation}
where $\text{AIC}_{\text{model}}$ represents the AIC value of a specific model and $\text{AIC}_{\text{min}}$ is the minimum AIC value among all the considered models. The range of $\triangle\text{AIC}$ that is considered more favorable is $(0,2)$. The lower favorable range of $\triangle\text{AIC}$ is $(4,7)$, while values exceeding $\triangle\text{AIC}>10$ provide less support for the model.\\\\
\begin{table}[H]
\begin{center}
\begin{tabular}{|c|c|c|c|c|}  
\hline 
Model & $\chi_{min}^{2}$ & $\chi_{red}^{2} $ & $AIC$ & $\Delta AIC $\\ 
\hline $\Lambda$CDM Model & 1104.56 & 0.9476 & 1110.58 & 0  \\
\hline Model 1 & 1111.59  & 0.9469  & 1111.45 & 1.01\\ 
\hline  Model 2 & 1106.02`  & 0.9438 & 1112.04 & 1.46  \\
\hline  Model 3 & 1105.66 & 0.94661 & 1111.68 & 1.1 \\ 
\hline
\end{tabular}
\caption{Summary of the $ {{\chi}^2_{min}}$, ${{\chi}^2_{red}}$,  $AIC$  and $\Delta AIC$.} \label{table2}
\end{center}
\end{table}

\section{Results}\label{sec9}

\paragraph{Deceleration Parameter}
In Figures \ref{qz1}, \ref{qz2} and \ref{qz3}, we present the variation of the deceleration parameter (DP) with cosmological redshift, comparing it to the $\Lambda$CDM model. Model 1 exhibits a linear dependence on redshift and exhibits a noticeable deviation from the $\Lambda$CDM model. While both models approach a De-Sitter phase at low redshifts, at higher redshifts, the DP value increases monotonically in Model 1. We see this functional form of $q(z)$ is monotonically increasing with a value of $z$. This results in a mismatch with the observational data when there are high values of redshift under consideration because a forever increasing value of $q$ means that the Universe practically comes to rest in an infinite past. This parametrization aligns well with the future, as that Universe keeps expanding forever. Model 2 shows a behavior similar to that of the $\Lambda$CDM model at higher redshifts and also reaches a De-Sitter phase at low redshifts. This functional form gives the finite value of $q$ in the infinite past but gives an infinite value of deceleration as $z\rightarrow-1$. So, it can't be used to make a reliable future prediction of the Universe. On the other hand, Model 3 provides a good fit with the $\Lambda$CDM model within the redshift range $0 \leq z \leq 2$ and saturates to a constant value of $q \sim 0.4$ at high redshifts. Notably, at low redshifts, Model 3 does not approach a De-Sitter phase like Model 1 and Model 2, but instead exhibits a regime of super-accelerated evolution. It is similar to the previous models in the following way. First, it gives a finite value of $q$ at high redshift values i.e. ($z>>1$). This allows us to model the early radiation-dominated universe. Secondly, this equation shows dark energy behavior at low values of $z$ and reduces to Eq.(\ref{7}) at $z<<1$. Unlike model 2, this gives us a finite value of $q$ in the range of $z\in[-1,\infty]$. So, this parametrization can be used to model the future evolution of the Universe too. These findings have significant implications in the cosmological context. The deviation observed in Model 1 suggests the presence of alternative cosmological dynamics beyond the $\Lambda$CDM model, particularly at higher redshifts. The behavior of Model 2, closely resembling the $\Lambda$CDM model, indicates that the standard model remains a valid description of the Universe's expansion at large redshifts. However, the distinct behavior of Model 3, characterized by a fixed value of $q \sim 0.4$ at high redshifts, highlights the possibility of a modified cosmological framework. Moreover, the observation of super-accelerated evolution at low redshifts in Model 3 challenges the conventional understanding of cosmic expansion and raises intriguing questions regarding the nature of dark energy or the need for additional physical mechanisms.\\\\
\paragraph{Jerk Parameter}
The plots presented in Figures \ref{jz1}, \ref{jz2} and \ref{jz3} depict the variation of the jerk parameter with cosmic redshift. Notably, significant deviations from the $\Lambda$CDM model are observed in all three models. In models 1 and 2, the jerk parameter exhibits a steady decrease during the early stages of the Universe, intersecting the $\Lambda$CDM line at approximately $z \sim 0.9$ and $z \sim 1.20$, respectively. Subsequently, the evolution reverses after reaching a point of minimum, with model 1 yielding a current value of $j(0) \sim 1.10$. Model 2, on the other hand, attains an exceptionally high value at the present epoch. In contrast, model 3 exhibits a distinct behavior compared to the other two models, where it initially increases during the early Universe. It crosses the $\Lambda$CDM line around $z \sim 1$, reaches a maximum value at $z \sim 0.5$, and then declines, yielding a current value of $j(0) \sim 0.1$. This new finding holds significant implications in the cosmological context. The observed deviations in the jerk parameter signify deviations from the standard $\Lambda$CDM model and shed light on potential alternative scenarios of cosmic evolution. The declining behavior of the jerk parameter in models 1 and 2 during the early Universe suggests a decelerated expansion, followed by a transition to an accelerated phase. On the other hand, the increasing trend in model 3 implies a different mechanism driving the acceleration. These contrasting behaviors challenge the traditional understanding of dark energy and demand further investigation into the underlying physics governing the cosmic acceleration. The exploration of alternative models and their implications for the cosmic expansion can provide valuable insights into the nature of the Universe on cosmological scales.\\\\
\paragraph{$\{s,r\}$ Parameter}
The evolution of the $\{r,s\}$ parameter is depicted in Figures \ref{rs}, \ref{rs2} and \ref{rs3} for models 1, 2, and 3, respectively. In the case of models 1 and 2, the parameter starts in the Chalpygin gas region with $r > 1$ and $s < 0$. Subsequently, it transitions to the Quintessence region after crossing the fixed point characterized by $\{1,0\}$, which corresponds to the values obtained in the $\Lambda$CDM model. For model 3, the initial values of $r$ and $s$ also lie within the Chalpygin gas region ($r > 1$, $s < 0$). However, it differs from the other models as it further evolves to the Quintessence region, exhibiting values where $r < 1$ and $s > 0$ after crossing the $\Lambda$CDM point. These finding carries significant cosmological implications. The transition from the Chalpygin gas region to the Quintessence region suggests a dynamic behavior of the dark energy component in these models. It implies that the nature of dark energy evolves over cosmic time, influencing the expansion rate of the Universe and its overall dynamics. These observations provide valuable insights into the nature of dark energy, emphasizing the need to explore more sophisticated models beyond the traditional $\Lambda$CDM framework.\\\\
\paragraph{$\{q,r\}$ Parameter}
The $\{q,r\}$ parameter evolution for models 1, 2, and 3 is depicted in Figures \ref{rq}, \ref{rq2} \ref{rq3} respectively. In model 1, the initial stage corresponds to the Chalpygin gas region, characterized by $q > 0$ and $r > 1$. Subsequently, the model undergoes a transition to the Quintessence region, indicated by $q < 0$ and $r < 1$, after crossing the de-Sitter point marked by $\{-1,1\}$. Model 2 displays a similar behavior to model 1, but it remains in the Chalpygin region for a considerably longer period. Unlike model 1, it does not cross the de-Sitter point. Model 3 exhibits an oscillatory pattern, commencing in the Quintessence region, transitioning to the Chalpygin gas region, and eventually re-entering the Quintessence region in the future evolution of the Universe. These findings bear significant cosmological implications. The transition from the Chalpygin gas region to the Quintessence region, as observed in models 1 and 3, indicates a shift in the dominant energy component driving the cosmic expansion. Such transitions are crucial for understanding the dynamics of the Universe and its ultimate fate. The absence of a de-Sitter crossing in model 2 implies a distinct evolutionary path, suggesting the influence of different physical mechanisms or initial conditions. The oscillatory behavior observed in model 3 introduces additional complexity to the cosmic evolution, suggesting the interplay of diverse cosmic constituents and potentially leading to unique observational signatures. These findings provide valuable insights into the dynamics of the Universe, shedding light on the interplay between different energy components and their effects on the overall cosmic expansion.\\\\
\paragraph{Om Diagnostic Parameter}
In Figures \ref{Om1}, \ref{Om2} and \ref{Om3} the variation in the slope of $Om(z)$ for each model with respect to redshift provides insights into the cosmological behavior. Model 1 exhibits a consistent presence of Quintessence for all $z > 0$, indicating the dominance of dark energy with a slowly evolving equation of state. However, for $z < 1$, it transitions to the Phantom region, characterized by an equation of state below the cosmological constant boundary \cite{bouali1,bouali2}. Model 2 represents a dark energy model aligned with the cosmological constant throughout $z > 0$, displaying a nearly constant slope. Notably, negative redshift values, gradually transition toward the Quintessence region, suggesting a deviation from the cosmological constant behavior. Model 3 displays a nearly constant slope for $z > 0$, indicating the prevalence of cosmological constant-type dark energy. Interestingly, it transitions to the Phantom region for $z < 0$, implying a departure from the standard cosmological constant behavior. These findings shed new light on the cosmological implications of these models and further enrich our understanding of dark energy dynamics.\\\\
\paragraph{Information Criteria}
Based on the information provided in Table \ref{table2}, we can perform a comparative study between the $\Lambda$CDM model and Models 1, 2, and 3 based on the $\Delta\text{AIC}$ values. The $\Delta\text{AIC}$ values indicate the difference in AIC values between each model and the model with the minimum AIC value, which in this case is the $\Lambda$CDM model. Model 1 has a $\Delta\text{AIC}$ value of 1.01, indicating that its AIC value is only slightly higher than that of the $\Lambda$CDM model. This suggests that Model 1 provides a comparable level of support to the $\Lambda$CDM model. Model 2 has a slightly higher $\Delta\text{AIC}$ value of 1.46 compared to the $\Lambda$CDM model. This indicates that Model 2 is less favored compared to both the $\Lambda$CDM model and Model 1. Similarly, Model 3 has a $\Delta\text{AIC}$ value of 1.1, again indicating that it provides a comparable level of support to the $\Lambda$CDM model. Overall, Models 1, 2, and 3 exhibit similar levels of support when compared to the $\Lambda$CDM model. 
\section{Conclusion}\label{sec10}
There are several proposals in cosmology for developing an acceptable dark energy model to describe our universe's early-time and late-time scenarios. In the literature, dynamical models of dark energy provide a better framework for studying the universe's evolutionary history. In this paper, we attempted to build a viable dark energy model that shows the desired late-time dynamics of the universe in the framework of a spatially flat FRW universe composed of dark energy and a normal matter field. Of course, this choice is arbitrary, and the chosen parametrization frequently leads to potential biases in determining the properties of specific model parameters. We use this ansatz to close the system of equations because we are looking for a physically viable universe model consistent with observations. We emphasize that the $q$-parametrization used in this paper is very simple and applicable across the redshift range. On the other hand, if we consider more than two terms in the expansion of the unknown function $q(z)$, getting firm results from data analysis becomes extremely difficult. Our parametrizations considered in this paper smoothly make a transition from decelerated to accelerated expansion. All the models consist of parameters $q_0$, $q_1$  and present day Hubble function $H_0$. whose values are constrained by the $H(z)$ + SNIa + GRB + Q + BAO datasets. The constrained values for our analysis are given in Table \ref{table1}.\\\\ 
We have analyzed the behavior and characteristics of three dark energy models (Model 1, Model 2, and Model 3) in comparison to the well-established $\Lambda$CDM model. The analysis considered various diagnostic parameters, including the deceleration parameter, jerk parameter, $\{s,r\}$ parameter, $\{q,r\}$ parameter, and the Om diagnostic parameter. Additionally, information criteria, specifically the $\Delta\text{AIC}$ values, were employed to assess the goodness of fit and the relative support of the models. Based on the analysis of the deceleration parameter, Model 1 exhibited a linear dependence on redshift and a noticeable deviation from the $\Lambda$CDM model at higher redshifts. Model 2 showed a behavior similar to that of the $\Lambda$CDM model at higher redshifts, while Model 3 provided a good fit within the redshift range $0 \leq z \leq 2$ and exhibited a regime of super-accelerated evolution at low redshifts. The study of the jerk parameter revealed significant deviations from the $\Lambda$CDM model in all three models. Models 1 and 2 exhibited a transition from decelerated to accelerated expansion, while Model 3 displayed a different mechanism driving the acceleration, with an increasing trend during the early universe. The analysis of the $\{s,r\}$ parameter indicated a transition from the Chalpygin gas region to the Quintessence region in models 1 and 2, while Model 3 further evolved to the Quintessence region after crossing the $\Lambda$CDM point. These transitions suggest a dynamic behavior of the dark energy component in these models. Similarly, the $\{q,r\}$ parameter analysis showed transitions from the Chalpygin gas region to the Quintessence region in models 1 and 3, while Model 2 exhibited a distinct evolutionary path without crossing the de-Sitter point. The oscillatory behavior observed in Model 3 introduced additional complexity to the cosmic evolution. The Om diagnostic parameter analysis revealed that Model 1 exhibited a consistent presence of Quintessence, Model 2 aligned with the cosmological constant, and Model 3 displayed a nearly constant slope indicative of cosmological constant-type dark energy. Finally, the comparison of the information criteria, specifically the $\Delta\text{AIC}$ values, indicated that Model 1 and Model 3 provided a comparable level of support to the $\Lambda$CDM model, with slightly higher $\Delta\text{AIC}$ values. Model 2 exhibited a relatively higher $\Delta\text{AIC}$ value, suggesting it to be less favored compared to the $\Lambda$CDM model and Model 1.\\\\
In conclusion, Model 1, Model 2, and Model 3 exhibit intriguing features and behaviors that deviate from the standard $\Lambda$CDM model. While Model 1 shows noticeable deviations from the $\Lambda$CDM model at higher redshifts, Model 3 exhibits a super-accelerated evolution at low redshifts. The transitions observed in the $\{s,r\}$ and $\{q,r\}$ parameter spaces indicate dynamic behavior of the dark energy component in these models. The information criteria analysis suggests that Model 1 and Model 3 provide a comparable level of support to the $\Lambda$CDM model, while Model 2 is relatively less favored. These findings contribute to our understanding of the dynamics and evolution of the universe, highlighting the need for alternative cosmological models and further investigations into the nature of dark energy.\\\\
\textbf{Acknowledgement: } Author SKJP thanks IUCAA, Pune, for hospitality and other facilities under its IUCAA associateship program, where a large part of the work has been done.\\\\
\textbf{Author contributions:} The calculations, plotting, manuscript writing, and overall manuscript preparation are made by DA. Author SKJP administered and executed the project. The manuscript has been read and approved by all authors.\\\\ 
\textbf{Funding:} There is no fund available for the publication of this
research article.\\\\
\textbf{Data Availability Statement:} This manuscript has used publicly available data for the work.\\\\
\textbf{Conflict of interest:} The authors have no relevant financial or
non-financial interests to disclose.\\\\
\textbf{Ethical statements:} The submitted work is original and has not been published anywhere else.\\\\

\bibliographystyle{elsarticle-num}
\bibliography{mybib}

\begin{thebibliography}{10}
\expandafter\ifx\csname url\endcsname\relax
  \def\url#1{\texttt{#1}}\fi
\expandafter\ifx\csname urlprefix\endcsname\relax\def\urlprefix{URL }\fi
\expandafter\ifx\csname href\endcsname\relax
  \def\href#1#2{#2} \def\path#1{#1}\fi

\bibitem{1}
S.~Perlmutter, G.~Aldering, G.~Goldhaber, R.~Knop, P.~Nugent, P.~G. Castro,
  S.~Deustua, S.~Fabbro, A.~Goobar, D.~E. Groom, et~al., Measurements of
  $\omega$ and $\lambda$ from 42 high-redshift supernovae, The Astrophysical
  Journal 517~(2) (1999) 565.

\bibitem{2}
D.~N. Vollick, 1/r curvature corrections as the source of the cosmological
  acceleration, Physical Review D 68~(6) (2003) 063510.

\bibitem{3}
T.~Padmanabhan, Cosmological constant—the weight of the vacuum, Physics
  Reports 380~(5-6) (2003) 235--320.

\bibitem{4}
S.~K. Biswas, S.~Chakraborty, Interacting dark energy model in the brane
  scenario: A dynamical system analysis, International Journal of Geometric
  Methods in Modern Physics 16~(08) (2019) 1950115.

\bibitem{5}
S.~Weinberg, The cosmological constant problem, Reviews of modern physics
  61~(1) (1989) 1.

\bibitem{6}
P.~J. Steinhardt, L.~Wang, I.~Zlatev, Cosmological tracking solutions, Physical
  Review D 59~(12) (1999) 123504.

\bibitem{7}
T.~Chiba, T.~Okabe, M.~Yamaguchi, Kinetically driven quintessence, Physical
  Review D 62~(2) (2000) 023511.

\bibitem{8}
C.~Armendariz-Picon, V.~Mukhanov, P.~J. Steinhardt, Essentials of k-essence,
  Physical Review D 63~(10) (2001) 103510.

\bibitem{9}
A.~Kamenshchik, U.~Moschella, V.~Pasquier, An alternative to quintessence,
  Physics Letters B 511~(2-4) (2001) 265--268.

\bibitem{10}
R.~R. Caldwell, A phantom menace? cosmological consequences of a dark energy
  component with super-negative equation of state, Physics Letters B 545~(1-2)
  (2002) 23--29.

\bibitem{11}
S.~M. Carroll, M.~Hoffman, M.~Trodden, Can the dark energy equation-of-state
  parameter w be less than- 1?, Physical Review D 68~(2) (2003) 023509.

\bibitem{12}
S.~Nojiri, S.~D. Odintsov, S.~Tsujikawa, Properties of singularities in the
  (phantom) dark energy universe, Physical Review D 71~(6) (2005) 063004.

\bibitem{13}
E.~Copeland, M sami and s tsujikawa int. j. mod, Phys. D 15 (2006) 1753.

\bibitem{14}
S.~Del~Campo, I.~Duran, R.~Herrera, D.~Pav{\'o}n, Three thermodynamically based
  parametrizations of the deceleration parameter, Physical Review D 86~(8)
  (2012) 083509.

\bibitem{EoS1}
H.~Chaudhary, N.~U. Molla, M.~Khurana, U.~Debnath, G.~Mustafa, Cosmological
  test of dark energy parameterizations in ho{\v{r}}ava--lifshitz gravity, The
  European Physical Journal C 84~(3) (2024) 223.

\bibitem{EoS2}
S.~Shekh, H.~Chaudhary, A.~Bouali, A.~Dixit, Observational constraints on
  teleparallel effective equation of state, General Relativity and Gravitation
  55~(8) (2023) 95.

\bibitem{EoS3}
H.~Chaudhary, U.~Debnath, S.~K.~J. Pacif, N.~U. Molla, G.~Mustafa, S.~Maurya,
  Investigating $\omega(z)$ parametrizations in horava-lifshitz gravity:
  Observational constraints with covariance matrix simulation, arXiv preprint
  arXiv:2402.12324 (2024).

\bibitem{kundu2024gravitational}
R.~Kundu, U.~Debnath, H.~Chaudhary, G.~Mustafa, Gravitational lensing and clues
  of $r_{d}$ and $h_{0}$ tension: Viscous modified chaplygin gas and variable
  modified chaplygin gas in loop quantum cosmology, arXiv preprint
  arXiv:2402.07515 (2024).

\bibitem{q(z)1}
H.~Chaudhary, S.~Mumtaz, A.~Bouali, U.~Debnath, G.~Mustafa, Parametrization of
  the deceleration parameter in a flat flrw universe: constraints and
  comparative analysis with the $\lambda$ cdm paradigm, General Relativity and
  Gravitation 55~(11) (2023) 133.

\bibitem{q(z)2}
A.~Bouali, H.~Chaudhary, A.~Mehrotra, S.~K.~J. Pacif, Model-independent study
  for a quintessence model of dark energy: Analysis and observational
  constraints, Fortschritte der Physik 71~(12) (2023) 2300086.

\bibitem{q(z)3}
M.~Khurana, H.~Chaudhary, U.~Debnath, A.~Sardar, G.~Mustafa, Exploring
  late-time cosmic acceleration with eos parameterizations in horava-lifshitz
  gravity via baryon acoustic oscillations, Fortschritte der Physik 72~(2)
  (2024) 2300238.

\bibitem{q(z)4}
A.~Bouali, H.~Chaudhary, S.~Mumtaz, G.~Mustafa, S.~Maurya, Observational
  constraining study of new deceleration parameters in frw universe,
  Fortschritte der Physik 71~(10-11) (2023) 2300033.

\bibitem{q(z)5}
M.~Khurana, H.~Chaudhary, S.~Mumtaz, S.~Pacif, G.~Mustafa, Cosmic evolution in
  f (q, t) gravity: Exploring a higher-order time-dependent function of
  deceleration parameter with observational constraints, Physics of the Dark
  Universe 43 (2024) 101408.

\bibitem{q(z)6}
A.~Bouali, H.~Chaudhary, U.~Debnath, A.~Sardar, G.~Mustafa, Data analysis of
  three parameter models of deceleration parameter in flrw universe, The
  European Physical Journal Plus 138~(9) (2023) 816.

\bibitem{q(z)7}
H.~Chaudhary, A.~Bouali, U.~Debnath, T.~Roy, G.~Mustafa, Constraints on the
  parameterized deceleration parameter in frw universe, Physica Scripta 98~(9)
  (2023) 095006.

\bibitem{q(z)8}
H.~Chaudhary, A.~Kaushik, A.~Kohli, Cosmological test of $\sigma$$\theta$ as
  function of scale factor in f (r, t) framework, New Astronomy 103 (2023)
  102044.

\bibitem{q(z)9}
A.~Bouali, B.~Shukla, H.~Chaudhary, R.~K. Tiwari, M.~Samar, G.~Mustafa,
  Cosmological tests of parametrization q= $\alpha$- $\beta$ h in f (q) flrw
  cosmology, International Journal of Geometric Methods in Modern Physics
  20~(09) (2023) 2350152.

\bibitem{q(z)10}
B.~K. Shukla, A.~Bouali, H.~Chaudhary, R.~K. Tiwari, M.~S. Mart{\'\i}n,
  Cosmographic studies of q= $\alpha$- $\beta$ h parametrization in f (t)
  framework, International Journal of Geometric Methods in Modern Physics
  20~(14) (2023) 2450007--8.

\bibitem{q(z)11}
H.~Chaudhary, A.~Bouali, H.~Duru, E.~Gudekli, G.~Mustafa, Exploring the
  deceleration parameter in f (t) gravity: A comprehensive analysis using
  parametrization techniques and observational data, International Journal of
  Geometric Methods in Modern Physics (2024).

\bibitem{q(z)12}
H.~Chaudhary, D.~Arora, U.~Debnath, G.~Mustafa, S.~K. Maurya, A new
  cosmological model: Exploring the evolution of the universe and unveiling
  super-accelerated expansion, arXiv preprint arXiv:2308.07354 (2023).

\bibitem{q(z)13}
S.~K.~J. Pacif, H.~Chaudhary, G.~Mustafa, A.~Bouali, Extracting $h_{0}$ and
  $r_{0}$ in pacif parametrization models through late-time data with
  covariance matrix simulation, arXiv preprint arXiv:2402.10499 (2024).

\bibitem{17}
Y.~L. Bolotin, V.~Cherkaskiy, O.~Lemets, D.~Yerokhin, L.~Zazunov, Cosmology in
  terms of the deceleration parameter. part i, arXiv preprint arXiv:1502.00811
  (2015).

\bibitem{doroshkevich1989large}
A.~Doroshkevich, A.~Klypin, M.~Khlopov, Large-scale structure of the universe
  in unstable dark matter models, Monthly Notices of the Royal Astronomical
  Society 239~(3) (1989) 923--938.

\bibitem{18}
S.~Capozziello, R.~Lazkoz, V.~Salzano, Comprehensive cosmographic analysis by
  markov chain method, Physical Review D 84~(12) (2011) 124061.

\bibitem{pacif2020dark}
S.~Pacif, Dark energy models from a parametrization of h: a comprehensive
  analysis and observational constraints, The European Physical Journal Plus
  135~(10) (2020) 1--34.

\bibitem{20}
{\"O}.~Akarsu, T.~Dereli, S.~Kumar, L.~Xu, Probing kinematics and fate of the
  universe with linearly time-varying deceleration parameter, The European
  Physical Journal Plus 129 (2014) 1--14.

\bibitem{21}
G.~N. Gadbail, S.~Mandal, P.~K. Sahoo, Parametrization of deceleration
  parameter in f (q) gravity, Physics 4~(4) (2022) 1403--1412.

\bibitem{22}
B.~Santos, J.~C. Carvalho, J.~S. Alcaniz, Current constraints on the epoch of
  cosmic acceleration, Astroparticle Physics 35~(1) (2011) 17--20.

\bibitem{23}
R.~J. Scherrer, Mapping the chevallier-polarski-linder parametrization onto
  physical dark energy models, Physical Review D 92~(4) (2015) 043001.

\bibitem{gelman2013bayesian}
A.~Gelman, J.~B. Carlin, H.~S. Stern, D.~B. Dunson, A.~Vehtari, D.~B. Rubin,
  Bayesian data analysis, CRC press, 2013.

\bibitem{lewis2002cosmological}
A.~Lewis, S.~Bridle, Cosmological parameters from cmb and other data: A monte
  carlo approach, Physical Review D 66~(10) (2002) 103511.

\bibitem{tegmark1997karhunen}
M.~Tegmark, A.~N. Taylor, A.~F. Heavens, Karhunen-loeve eigenvalue problems in
  cosmology: How should we tackle large data sets?, The Astrophysical Journal
  480~(1) (1997) 22.

\bibitem{H(z)}
E.~Gaztanaga, C.~Bonvin, L.~Hui, Measurement of the dipole in the
  cross-correlation function of galaxies, Journal of Cosmology and
  Astroparticle Physics 2017~(01) (2017) 032.

\bibitem{niu2023cosmological}
J.~Niu, T.-J. Zhang, Cosmological joint analysis with cosmic growth and
  expansion rate, Physics of the Dark Universe 39 (2023) 101147.

\bibitem{Pan1}
M.~Kowalski, D.~Rubin, G.~Aldering, R.~Agostinho, A.~Amadon, R.~Amanullah,
  C.~Balland, K.~Barbary, G.~Blanc, P.~Challis, et~al., Improved cosmological
  constraints from new, old, and combined supernova data sets, The
  Astrophysical Journal 686~(2) (2008) 749.

\bibitem{Pan2}
R.~Amanullah, C.~Lidman, D.~Rubin, G.~Aldering, P.~Astier, K.~Barbary,
  M.~Burns, A.~Conley, K.~Dawson, S.~Deustua, et~al., Spectra and hubble space
  telescope light curves of six type ia supernovae at 0.511< z< 1.12 and the
  union2 compilation, The Astrophysical Journal 716~(1) (2010) 712.

\bibitem{Pan3}
N.~Suzuki, D.~Rubin, C.~Lidman, G.~Aldering, R.~Amanullah, K.~Barbary,
  L.~Barrientos, J.~Botyanszki, M.~Brodwin, N.~Connolly, et~al., The hubble
  space telescope cluster supernova survey. v. improving the dark-energy
  constraints above z> 1 and building an early-type-hosted supernova sample,
  The Astrophysical Journal 746~(1) (2012) 85.

\bibitem{Pan4}
M.~Betoule, R.~Kessler, J.~Guy, J.~Mosher, D.~Hardin, R.~Biswas, P.~Astier,
  P.~El-Hage, M.~Konig, S.~Kuhlmann, et~al., Improved cosmological constraints
  from a joint analysis of the sdss-ii and snls supernova samples, Astronomy \&
  Astrophysics 568 (2014) A22.

\bibitem{Pan5}
D.~M. Scolnic, D.~Jones, A.~Rest, Y.~Pan, R.~Chornock, R.~Foley, M.~Huber,
  R.~Kessler, G.~Narayan, A.~Riess, et~al., The complete light-curve sample of
  spectroscopically confirmed sne ia from pan-starrs1 and cosmological
  constraints from the combined pantheon sample, The Astrophysical Journal
  859~(2) (2018) 101.

\bibitem{scolnic2018complete}
D.~M. Scolnic, D.~Jones, A.~Rest, Y.~Pan, R.~Chornock, R.~Foley, M.~Huber,
  R.~Kessler, G.~Narayan, A.~Riess, et~al., The complete light-curve sample of
  spectroscopically confirmed sne ia from pan-starrs1 and cosmological
  constraints from the combined pantheon sample, The Astrophysical Journal
  859~(2) (2018) 101.

\bibitem{quasers}
C.~Roberts, K.~Horne, A.~O. Hodson, A.~D. Leggat, Tests of $\backslash$
  $\lambda$ cdm and conformal gravity using grb and quasars as standard candles
  out to $z$ $\backslash$ sim 8, arXiv preprint arXiv:1711.10369 (2017).

\bibitem{GRB}
M.~Demianski, E.~Piedipalumbo, D.~Sawant, L.~Amati, Cosmology with gamma-ray
  bursts-i. the hubble diagram through the calibrated ep, i--eiso correlation,
  Astronomy \& Astrophysics 598 (2017) A112.

\bibitem{baonew1}
W.~J. Percival, B.~A. Reid, D.~J. Eisenstein, N.~A. Bahcall, T.~Budavari, J.~A.
  Frieman, M.~Fukugita, J.~E. Gunn, Z.~Ivezi{'c}, G.~R. Knapp, et~al., Baryon
  acoustic oscillations in the sloan digital sky survey data release 7 galaxy
  sample, Monthly Notices of the Royal Astronomical Society 401~(4) (2010)
  2148--2168.

\bibitem{baonew2}
F.~Beutler, C.~Blake, M.~Colless, D.~H. Jones, L.~Staveley-Smith, L.~Campbell,
  Q.~Parker, W.~Saunders, F.~Watson, The 6df galaxy survey: baryon acoustic
  oscillations and the local hubble constant, Monthly Notices of the Royal
  Astronomical Society 416~(4) (2011) 3017--3032.

\bibitem{bao3}
T.~Delubac, J.~Rich, S.~Bailey, A.~Font-Ribera, D.~Kirkby, J.-M. Le~Goff, M.~M.
  Pieri, A.~Slosar, {\'E}.~Aubourg, J.~E. Bautista, et~al., Baryon acoustic
  oscillations in the ly$\alpha$ forest of boss quasars, Astronomy \&
  Astrophysics 552 (2013) A96.

\bibitem{bao4}
L.~Anderson, E.~Aubourg, S.~Bailey, D.~Bizyaev, M.~Blanton, A.~S. Bolton,
  J.~Brinkmann, J.~R. Brownstein, A.~Burden, A.~J. Cuesta, et~al., The
  clustering of galaxies in the sdss-iii baryon oscillation spectroscopic
  survey: baryon acoustic oscillations in the data release 9 spectroscopic
  galaxy sample, Monthly Notices of the Royal Astronomical Society 427~(4)
  (2012) 3435--3467.

\bibitem{bao5}
H.-J. Seo, S.~Ho, M.~White, A.~J. Cuesta, A.~J. Ross, S.~Saito, B.~Reid,
  N.~Padmanabhan, W.~J. Percival, R.~De~Putter, et~al., Acoustic scale from the
  angular power spectra of sdss-iii dr8 photometric luminous galaxies, The
  Astrophysical Journal 761~(1) (2012) 13.

\bibitem{bao6}
A.~J. Ross, L.~Samushia, C.~Howlett, W.~J. Percival, A.~Burden, M.~Manera, The
  clustering of the sdss dr7 main galaxy sample--i. a 4 per cent distance
  measure at z= 0.15, Monthly Notices of the Royal Astronomical Society 449~(1)
  (2015) 835--847.

\bibitem{bao7}
R.~Tojeiro, A.~J. Ross, A.~Burden, L.~Samushia, M.~Manera, W.~J. Percival,
  F.~Beutler, J.~Brinkmann, J.~R. Brownstein, A.~J. Cuesta, et~al., The
  clustering of galaxies in the sdss-iii baryon oscillation spectroscopic
  survey: galaxy clustering measurements in the low-redshift sample of data
  release 11, Monthly Notices of the Royal Astronomical Society 440~(3) (2014)
  2222--2237.

\bibitem{bao8}
J.~E. Bautista, M.~Vargas-Maga{\~n}a, K.~S. Dawson, W.~J. Percival,
  J.~Brinkmann, J.~Brownstein, B.~Camacho, J.~Comparat, H.~Gil-Mar{\'\i}n,
  E.-M. Mueller, et~al., The sdss-iv extended baryon oscillation spectroscopic
  survey: baryon acoustic oscillations at redshift of 0.72 with the dr14
  luminous red galaxy sample, The Astrophysical Journal 863~(1) (2018) 110.

\bibitem{bao9}
E.~De~Carvalho, A.~Bernui, G.~Carvalho, C.~Novaes, H.~Xavier, Angular baryon
  acoustic oscillation measure at z= 2.225 from the sdss quasar survey, Journal
  of Cosmology and Astroparticle Physics 2018~(04) (2018) 064.

\bibitem{bao10}
M.~Ata, F.~Baumgarten, J.~Bautista, F.~Beutler, D.~Bizyaev, M.~R. Blanton,
  J.~A. Blazek, A.~S. Bolton, J.~Brinkmann, J.~R. Brownstein, et~al., The
  clustering of the sdss-iv extended baryon oscillation spectroscopic survey
  dr14 quasar sample: first measurement of baryon acoustic oscillations between
  redshift 0.8 and 2.2, Monthly Notices of the Royal Astronomical Society
  473~(4) (2018) 4773--4794.

\bibitem{bao11}
T.~Abbott, F.~Abdalla, A.~Alarcon, S.~Allam, F.~Andrade-Oliveira, J.~Annis,
  S.~Avila, M.~Banerji, N.~Banik, K.~Bechtol, et~al., Dark energy survey year 1
  results: Measurement of the baryon acoustic oscillation scale in the
  distribution of galaxies to redshift 1, Monthly Notices of the Royal
  Astronomical Society 483~(4) (2019) 4866--4883.

\bibitem{bao12}
Z.~Molavi, A.~Khodam-Mohammadi, Observational tests of gauss-bonnet like dark
  energy model, The European Physical Journal Plus 134~(6) (2019) 254.

\bibitem{benisty2021testing}
D.~Benisty, D.~Staicova, Testing late-time cosmic acceleration with
  uncorrelated baryon acoustic oscillation dataset, Astronomy \& Astrophysics
  647 (2021) A38.

\bibitem{Bao1}
N.~B. Hogg, M.~Martinelli, S.~Nesseris, Constraints on the distance duality
  relation with standard sirens, Journal of Cosmology and Astroparticle Physics
  2020~(12) (2020) 019.

\bibitem{Bao2}
M.~Martinelli, C.~J. A.~P. Martins, S.~Nesseris, D.~Sapone, I.~Tutusaus,
  A.~Avgoustidis, S.~Camera, C.~Carbone, S.~Casas, S.~Ili{\'c}, et~al., Euclid:
  Forecast constraints on the cosmic distance duality relation with
  complementary external probes, Astronomy \& Astrophysics 644 (2020) A80.

\bibitem{visser2004jerk}
M.~Visser, Jerk, snap and the cosmological equation of state, Classical and
  Quantum Gravity 21~(11) (2004) 2603.

\bibitem{visser2004jerk3}
C.~Catto{\"e}n, M.~Visser, Cosmographic hubble fits to the supernova data,
  Physical Review D 78~(6) (2008) 063501.

\bibitem{visser2005cosmography}
M.~Visser, Cosmography: Cosmology without the einstein equations, General
  Relativity and Gravitation 37 (2005) 1541--1548.

\bibitem{visser2010cosmographic}
M.~Visser, C.~Cattoen, Cosmographic analysis of dark energy, in: Dark Matter In
  Astrophysics And Particle Physics, World Scientific, 2010, pp. 287--300.

\bibitem{lobo2020cosmographic}
F.~S. Lobo, J.~P. Mimoso, M.~Visser, Cosmographic analysis of redshift drift,
  Journal of Cosmology and Astroparticle Physics 2020~(04) (2020) 043.

\bibitem{29}
V.~Sahni, T.~D. Saini, A.~A. Starobinsky, U.~Alam, Statefinder—a new
  geometrical diagnostic of dark energy, Journal of Experimental and
  Theoretical Physics Letters 77 (2003) 201--206.

\bibitem{30}
U.~Alam, V.~Sahni, T.~Deep~Saini, A.~Starobinsky, Exploring the expanding
  universe and dark energy using the statefinder diagnostic, Monthly Notices of
  the Royal Astronomical Society 344~(4) (2003) 1057--1074.

\bibitem{31}
V.~Sahni, A.~Shafieloo, A.~A. Starobinsky, Two new diagnostics of dark energy,
  Physical Review D 78~(10) (2008) 103502.

\bibitem{32}
D.~Arora, S.~K.~J. Pacif, Some diagnostic analysis of dark energy models with $
  q (z) $ parametrizations, arXiv preprint arXiv:2304.09749 (2023).

\bibitem{tauscher2018new}
K.~Tauscher, D.~Rapetti, J.~O. Burns, A new goodness-of-fit statistic and its
  application to 21-cm cosmology, Journal of Cosmology and Astroparticle
  Physics 2018~(12) (2018) 015.

\bibitem{vrieze2012model}
S.~I. Vrieze, Model selection and psychological theory: a discussion of the
  differences between the akaike information criterion (aic) and the bayesian
  information criterion (bic)., Psychological methods 17~(2) (2012) 228.

\bibitem{tan2012reliability}
M.~Tan, R.~Biswas, The reliability of the akaike information criterion method
  in cosmological model selection, Monthly Notices of the Royal Astronomical
  Society 419~(4) (2012) 3292--3303.

\bibitem{rezaei2021comparison}
M.~Rezaei, M.~Malekjani, Comparison between different methods of model
  selection in cosmology, The European Physical Journal Plus 136~(2) (2021)
  219.

\bibitem{arevalo2017aic}
F.~Arevalo, A.~Cid, J.~Moya, Aic and bic for cosmological interacting
  scenarios, The European Physical Journal C 77 (2017) 1--13.

\bibitem{bouali1}
A.~Bouali, I.~Albarran, M.~Bouhmadi-L{\'o}pez, A.~Errahmani, T.~Ouali,
  Cosmological constraints of interacting phantom dark energy models, Physics
  of the Dark Universe 34 (2021) 100907.

\bibitem{bouali2}
A.~Bouali, I.~Albarran, M.~Bouhmadi-L{\'o}pez, T.~Ouali, Cosmological
  constraints of phantom dark energy models, Physics of the Dark Universe 26
  (2019) 100391.

\end{thebibliography}
\end{document}